\documentclass[notitlepage,a4paper,aps,prd,tightenlines,preprintnumbers,nofootinbib,showkeys,superscriptaddress,9pt,longbibliography]{revtex4-1}
\pdfoutput=1

\usepackage{listings}
\usepackage[scaled=0.95]{XCharter}
\usepackage[T1]{fontenc}
\usepackage{pifont}
\usepackage[utf8]{inputenc}
\usepackage{inconsolata}

\usepackage{hhline}
\usepackage{mathptmx}
\usepackage{diagbox}
\usepackage{fullpage}
\usepackage{amsfonts}
\usepackage{amsmath}
\usepackage{slashed}
\usepackage{tabularx}
\usepackage{amssymb}
\usepackage{graphicx}
\usepackage{array}
\usepackage{epic}
\usepackage{rotating}
\usepackage{eepic}
\usepackage{epsfig}
\usepackage{latexsym}
\usepackage{booktabs}
\usepackage[dvipsnames]{xcolor}
\usepackage[export]{adjustbox}
\usepackage{colortbl}
\usepackage{xcolor}
\usepackage{float}
\usepackage{multirow}
\usepackage[linktocpage]{hyperref}
\usepackage{enumitem}
\hypersetup{colorlinks=true,citecolor=brown,linkcolor=Brown,urlcolor=NavyBlue}
\usepackage[caption=false]{subfig}
 
\usepackage{natbib}

\usepackage{relsize}
\usepackage[left=2.4cm,right=2.4cm,top=2.0cm,bottom=2.5cm]{geometry}
\usepackage{listings}

\usepackage{xltabular}

\definecolor{OliveGreen}{rgb}{0,0.6,0}
\usepackage{shorthand}

\newsavebox\solbox
\makeatletter
{
\@parboxrestore%
\begin{lrbox}{\solbox}%
\begin{minipage}{0.8\linewidth}
}
{\end{minipage}\end{lrbox}%
\begin{center}
\framebox{\usebox\solbox}
\end{center}
}
\makeatother

\begin{document}

\title{\textsc{TooLQit}: Leptoquark Models and Limits}

\author{Arvind Bhaskar}
\email{arvind.bhaskar@iopb.res.in}
\affiliation{Institute of Physics, Sachivalaya Marg, Bhubaneswar 751 005, India}
\affiliation{Center for Computational Natural Sciences and Bioinformatics, International Institute of Information Technology, Hyderabad 500 032, India}

\author{Yash Chaurasia}
\email{yash.chaurasia@research.iiit.ac.in}
\affiliation{Center for Computational Natural Sciences and Bioinformatics, International Institute of Information Technology, Hyderabad 500 032, India}

\author{Arijit Das}
\email{arijit21@iisertvm.ac.in}
\affiliation{Indian Institute of Science Education and Research Thiruvananthapuram, Vithura, Kerala, 695 551, India}

\author{Atirek Kumar}
\email{atirek.kumar@research.iiit.ac.in}
\affiliation{Center for Computational Natural Sciences and Bioinformatics, International Institute of Information Technology, Hyderabad 500 032, India}

\author{Tanumoy Mandal}
\email{tanumoy@iisertvm.ac.in}
\affiliation{Indian Institute of Science Education and Research Thiruvananthapuram, Vithura, Kerala, 695 551, India}

\author{Subhadip Mitra}
\email{subhadip.mitra@iiit.ac.in}
\affiliation{Center for Computational Natural Sciences and Bioinformatics, International Institute of Information Technology, Hyderabad 500 032, India}
\affiliation{Center for Quantum Science and Technology, International Institute of Information Technology, Hyderabad 500 032, India}

\author{Cyrin Neeraj}
\email{cyrin.neeraj@research.iiit.ac.in}
\affiliation{Center for Computational Natural Sciences and Bioinformatics, International Institute of Information Technology, Hyderabad 500 032, India}

\author{Rachit Sharma}
\email{rachit21@iisertvm.ac.in}
\affiliation{Indian Institute of Science Education and Research Thiruvananthapuram, Vithura, Kerala, 695 551, India}

\begin{abstract}
\noindent
We introduce the leptoquark (LQ) toolkit, \textsc{TooLQit}, which includes leading-order \textsc{FeynRules} models for all types of LQs and a \textsc{Python}-based calculator, named \textsc{CaLQ}, to test if a set of parameter points are allowed by the LHC dilepton searches. The models include electroweak gauge interactions of the LQs and follow a set of intuitive notations. Currently, \textsc{CaLQ} can calculate the LHC limits on LQ ($S_1$ and $U_1$) couplings (one or more simultaneously) for any mass between $1$ and $5$ TeV using a $\chi^2$ method. In this manual for \textsc{TooLQit}, we describe the \textsc{FeynRules} models and discuss the techniques used in \textsc{CaLQ}. We outline the workflow to check parameter spaces of LQ models with an example. We show some illustrative scans for one- and multi-coupling scenarios for the $U_1$ vector LQ. The \textsc{TooLQit} code is available at \url{https://github.com/rsrchtsm/TooLQit}.

\end{abstract}

\renewcommand\labelitemi{\large \ding{113}}
\maketitle 

\section{Introduction}
\label{sec:intro}
\noindent
Many beyond-the-Standard Model (BSM) scenarios (e.g., Pati-Salam model~\cite{Pati:1973uk,Pati:1974yy}, grand unified theories~\cite{Georgi:1974sy,Fritzsch:1974nn}, quark-lepton compositeness~\cite{Schrempp:1984nj,Gripaios:2009dq}, coloured Zee-Babu models~\cite{Kohda:2012sr}, technicolor models~\cite{Dimopoulos:1979es,Farhi:1980xs} and $R$-parity--violating supersymmetric models~\cite{Barbier:2004ez}, etc.) contain coloured scalar or vector bosons with nonzero lepton numbers in the TeV range. There are several possibilities for these scalar or vector particles---commonly called leptoquarks (LQs)~\cite{Buchmuller:1986zs,Blumlein:1994qd,Blumlein:1996qp,Dorsner:2016wpm}---based on their weak representations, which are well-studied in the literature. Recently, LQs attracted significant attention mainly in the context of various experimental anomalies like the one observed in the ratios of $B$-meson semileptonic decays ($R_{D^{(*)}}$, which still exhibit a combined $3.3\sigma$ deviation from theoretical predictions~\cite{HFLAV:2024ctg}) or the anomalous magnetic moment of the muon $(g-2)_{\mu}$~\cite{Aoyama:2020ynm}, etc. There are other theoretical/phenomenological motivations for TeV-scale LQs as well. For example, they can explain baryon asymmetry via leptogenesis~\cite{Fong:2013gaa}, enhance the production of colour neutral particles~\cite{Das:2017kkm,Bhaskar:2023xkm,Bigaran:2023ris}, play roles in Higgs physics~\cite{Cheung:2015yga,Bhaskar:2020kdr,Bhaskar:2022ygp}, can act as a portal to dark matter~\cite{Choi:2018stw,Mandal:2018czf}, can stabilise the electroweak vacuum~\cite{Bandyopadhyay:2016oif}, and have the potential to produce gravitational waves by inducing first-order electroweak phase transition~\cite{Fu:2022eun}, etc.

On the experimental side, these particles are well-explored. Since LQs simultaneously decay to quarks and leptons, their signatures are unique. The LHC experiments show good sensitivity towards most LQ models mainly because of the presence of leptons in the final states. Both CMS~\cite{CMS:2023plots} and ATLAS~\cite{ATLAS:2024yxs} collaborations have dedicated LQ search programs. They have extensively looked for signatures of LQ productions in various final states (like dilepton-dijet final states from LQ pair productions) and, so far, have put strong bounds on LQ parameters. The current mass exclusion bounds on scalar LQs from the pair production searches are within $1$-$2$ TeV; for vectors, the limits are stronger by a few hundred GeVs. 

Since, at the LHC, LQs are produced in pairs mainly via QCD interactions, these can be interpreted as model-independent lower bounds on LQ masses, assuming the unknown Yukawa couplings (LQ-quark-lepton) responsible for LQ decays are small enough not to affect their production cross-section~\cite{Diaz:2017lit}. However, these new couplings are not small in various BSM scenarios. Hence, the new couplings may also contribute to the production processes~\cite{Schmaltz:2018nls}. Large new couplings will contribute to the pair productions and open up new processes like single and indirect or non-resonant (i.e., $t$-channel LQ exchange and its interference with the SM background) productions~\cite{Belyaev:2005ew,Mandal:2015vfa,Mandal:2018kau}. The presence of LQs can also be inferred from other processes. For example, with order-one new coupling(s), TeV-range LQs will lead to observable shifts in the high-$p_T$ tails of the dilepton or lepton plus missing transverse energy distributions. Hence, depending on the sizes of the new couplings, the LQ models can have better prospects and stricter limits~\cite{Faroughy:2016osc,Greljo:2017vvb,Mandal:2018kau,Greljo:2018tzh,Baker:2019sli,Aydemir:2019ynb,Chandak:2019iwj,Angelescu:2020uug,Bhaskar:2020gkk,Bhaskar:2021pml,Bhaskar:2021gsy,Aydemir:2022lrq,Bhaskar:2023ftn,Bhaskar:2024swq}.

As the LHC records more and more data, its sensitivity towards BSM processes with small cross-sections increases with improving statistics. Many processes that were difficult (or impossible) to probe with the data collected in the earlier runs will come within its reach. Hence, with the ongoing Run III, we need to ensure that the BSM signal simulations are not (unnecessarily) neglecting contributions that could be within the reach of the LHC at high luminosity. One way to do that will be to simulate signals at higher orders. For example, currently, various BSM processes can be simulated at the next-to-leading order (NLO) in QCD~\cite{Degrande:2014vpa,Hirschi:2015iia,Degrande:2020evl} (see Refs.~\cite{Mandal:2015lca,Dorsner:2018ynv,Buonocore:2022msy,Korajac:2023xtv} for LQ processes in particular). However, there is another possible direction for improvement. Some electroweak contributions, which can lead to observable effects, are ignored in current simulations. For example, in Ref.~\cite{Bhaskar:2023ftn}, we showed how the oft-ignored photon/$Z$-mediated LQ productions lead to noticeable shifts in the (model-independent) exclusion limits. 
 
In this paper, we introduce \textsc{TooLQit}---a toolkit with \textsc{FeynRules}~\cite{Alloul:2013bka} models of all possible LQs coupling exclusively to the SM particles for Monte Carlo simulations and a calculator, \textsc{CaLQ}, to test the Yukawa couplings of LQs against the indirect limits from the current dilepton data. While the models are leading order (LO) at present, they include the electroweak vertices, including the photon/$Z$-gluon-LQ-LQ vertex (since LQs carry electric charges and are colour triplets, such a vertex is possible). For ease of use, we introduce a set of intuitive notations for the LQs and the new couplings. \textsc{CaLQ} is a {\sc Python} package to test if a parameter point (i.e., the mass of the LQ and a set of nonzero LQ-$q$-$\ell$ couplings) is allowed by the LHC indirect limits. Currently, it uses only the dilepton search data~\cite{Aad:2020zxo,Sirunyan:2021khd} and supports two weak-singlet LQs---the charge-$1/3$ scalar $S_1$ and the charge-$2/3$ vector $U_1$---for masses between $1$ and $5$ TeV. (Similar \textsc{Mathematica}-based LQ and SM effective field theory limit calculator \textsc{HighPT} utilises the high-$p_T$  dilepton and lepton $+\ \slashed{E}_T$ tails~\cite{Allwicher:2022gkm,Allwicher:2022mcg}. However, at present it calculates limits only for three LQ mediator masses: $1$, $2$ and $3$ TeV.) It works on a $\chi^2$ minimisation method we developed in Ref.~\cite{Mandal:2018kau} (also~\cite{Aydemir:2019ynb}), and generalised in Ref.~\cite{Bhaskar:2021pml}.

In Section~\ref{sec:models}, we explain our notations and describe the FeynRules models and in Section~\ref{sec:calq}, we describe the calculator.

\section{Leptoquark Models}\label{sec:models}
\noindent
As listed in Ref.~\cite{Dorsner:2016wpm},
there are twelve possible renormalizable LQ models: six scalars (commonly referred to as $S_1, \widetilde{S}_1, \overline{S}_1, R_2, \widetilde{R}_2$, $S_3$) and six vectors ($U_1, \widetilde{U}_1, \overline{U}_1, V_2, \widetilde{V}_2, U_3$). Their $SU(2)_{L}$ structures are (in the EM charge basis) shown below (the superscripts show the electric charges):
\begin{equation}
\def\arraystretch{1.3}
\begin{array}{@{}ccccc}
S_{1} \equiv
    \begin{pmatrix}
      S^{\frac{1}{3}}_{1} \\
    \end{pmatrix}, \widetilde{S}_{1}\equiv
    \begin{pmatrix}
      \widetilde{S}^{\frac{4}{3}}_{1} \\
    \end{pmatrix}, \widetilde{S}_{1}\equiv
    \begin{pmatrix}
      \widetilde{S}^{\frac{4}{3}}_{1} \\
    \end{pmatrix}
&~~~~~~~~&
R_{2} \equiv
    \begin{pmatrix}
      R^{\frac{5}{3}}_{2} \\
      R^{\frac{2}{3}}_{2} \\
    \end{pmatrix}, \widetilde{R}_{2}\equiv
    \begin{pmatrix}
      \widetilde{R}^{\frac{2}{3}}_{2} \\
      \widetilde{R}^{-\frac{1}{3}}_{2} \\
    \end{pmatrix}
&~~~~~~~~&
S_{3}\equiv
    \begin{pmatrix}
      S^{4/3}_{3} \\
      S^{1/3}_{3} \\
      S^{-2/3}_{3} \\
    \end{pmatrix}
\\~\\
U_{1}\equiv
    \begin{pmatrix}
      U^{\frac{2}{3}}_{1} \\
    \end{pmatrix}, \widetilde{U}_{1}\equiv
    \begin{pmatrix}
      \widetilde{U}^{\frac{5}{3}}_{1} 
    \end{pmatrix}, \overline{U}_1\equiv
    \begin{pmatrix}
      \overline{U}^{-\frac{1}{3}}_{1} \\
    \end{pmatrix}
&~~~~~~~~&
V_{2}\equiv
    \begin{pmatrix}
      V^{\frac{5}{3}}_{2} \\
      V^{\frac{2}{3}}_{2} \\
    \end{pmatrix}, \widetilde{V}_{2}\equiv
    \begin{pmatrix}
      \widetilde{V}^{\frac{2}{3}}_{2} \\
      \widetilde{V}^{-\frac{1}{3}}_{2} \\
    \end{pmatrix}
&~~~~~~~~&    
U_{3}\equiv
    \begin{pmatrix}
      U^{5/3}_{3} \\
      U^{2/3}_{3} \\
      U^{-1/3}_{3} \\
    \end{pmatrix}
\end{array}
\end{equation}
The hypercharges are obtained by the Gell-Mann-Nishijima formula: $Q = T^3 + Y$ where $T^3$ is the third component of the weak-isospin.\medskip

\begin{table*}
\caption{Yukawa interactions in up- and down-aligned scenarios for scalar and vector LQs excluding diquark interactions. Throughout the paper, a slightly modified and more explicit notation for the Yukawa couplings is adopted.
\label{tab:SLQYuk}}
\centering{\footnotesize\renewcommand\baselinestretch{2}\selectfont
\begin{tabular*}{\textwidth}{l @{\extracolsep{\fill}}cc}
\hline \hline LQ Model & Down-aligned Yukawa Interactions & Up-aligned Yukawa Interactions \\ \hline\hline
$S_{1}$     & $-y^{LL}_{1ij}\ \overline{d^c_L}^{i} \nu^{j}_{L}S_{1} + (V^{*}y_{1}^{LL})_{ij}\ \overline{u^c_L}^{i} e^{j}_{L}S_{1} + y^{RR}_{1ij}\ \overline{u^c_R}^{i}e^{j}_{R} S_{1}$            & 
$-(V^{T}y^{LL}_1)_{ij}\ \overline{d^c_L}^{i} \nu^{j}_{L}S_{1} + y^{LL}_{1ij}\ \overline{u^c_L}^{i} e^{j}_{L}S_{1} + y^{RR}_{1ij}\ \overline{u^c_R}^{i} e^{j}_{R}S_{1}$ \\ 
$\widetilde{S}_{1}$     & \multicolumn{2}{c}{$\widetilde{y}^{RR}_{1ij}\ \overline{d^c_{R}}^{i}e^{j}_{R}\widetilde{S}_{1}$}           \\ 

\multirow{2}{*}{$R_{2}$}     & $-y^{RL}_{2ij}\ (\overline{u}_{R}^{i}e^{j}_{L}R^{5/3}_{2} - \overline{u}_{R}^{i}\nu^{j}_{L}R^{2/3}_{2})$            & $-y^{RL}_{2ij}\ (\overline{u}_{R}^{i}e^{j}_{L}R^{5/3}_{2} - \overline{u}_{R}^{i}\nu^{j}_{L}R^{2/3}_{2})$          \\ 
&$ + (Vy_{2}^{LR})_{ij}\ \overline{u}_L^{i}e^{j}_{R}R^{5/3}_{2} + y^{{LR}}_{2ij}\ \overline{d}_L^{i}e^{j}_{R}R^{2/3}_{2}$&$ + y^{LR}_{2ij}\ \overline{u}_L^{i}e^{j}_{R}R^{5/3}_{2} + (V^{\dagger}y_{2}^{LR})_{ij}\ \overline{d}_L^{i}e^{j}_{R}R^{2/3}_{2}$\\
$\widetilde{R}_{2}$     & \multicolumn{2}{c}{$-\widetilde{y}^{RL}_{2ij}\ 
(\overline{d}^{i}_{R}e^{j}_{L}\widetilde{R}^{2/3}_{2} - 
\overline{d}^{i}_{R}\nu^{j}_{L}\widetilde{R}^{-1/3}_{2})$} \\ 
 \multirow{2}{*}{$S_{3}$}     & $-y^{LL}_{3ij}\ \overline{d^c_{L}}^{i}\nu^{j}_{L}S^{1/3}_{3}-(V^{*}y_{3}^{LL})_{ij}\ \overline{u^c_{L}}^{i}e^{j}_{L}S^{1/3}_{3}$             & $-(V^Ty_{3}^{LL})_{ij}\ \overline{d^c_{L}}^{i}\nu^{j}_{L}S^{1/3}_{3}-y^{LL}_{3ij}\ \overline{u^c_{L}}^{i}e^{j}_{L}S^{1/3}_{3}$         \\ 
 &  $-\sqrt{2}y^{LL}_{3ij}\ \overline{d^c_{L}}^{i}e^{j}_{L}S^{4/3}_{3}+\sqrt{2}(V^* y^{LL}_{3})_{ij}\ \overline{u^c_{L}}^{i}\nu^{j}_{L}S^{-2/3}_{3}$ & $-\sqrt{2}(V^T y_{3}^{LL})_{ij}\ \overline{d^c_{L}}^{i}e^{j}_{L}S^{4/3}_{3}+\sqrt{2}y^{LL}_{3ij}\  \overline{u^c_{L}}^{i}\nu^{j}_{L}S^{-2/3}_{3}$ \\
\multirow{2}{*}{$U_1$}     & $(Vx_{1}^{LL})_{1ij}\ \overline{u}_{L}^{i}\gamma^{\mu}\nu^{j}_{L}U_{1,\mu} + x_{1ij}^{LL} \overline{d}_{L}^{i}\gamma^{\mu}e^{j}_{L}U_{1,\mu}$            & $x^{LL}_{1ij}\ \overline{u}_{L}^{i}\gamma^{\mu}\nu^{j}_{L}U_{1,\mu} + (V^{\dagger}x^{LL}_{1})_{ij}\overline{d}_{L}^{i}\gamma_{\mu}e^{j}_{L}U_{1,\mu}$          \\ 
&$ + x^{RR}_{1ij}\ \overline{d}_R^{i}\gamma^{\mu}e^{j}_{R}U_{1,\mu} + x^{\overline{RR}}_{1ij}\ \overline{u}_R^{i}\gamma^{\mu}\nu^{j}_{R}U_{1,\mu}$&$ + x^{RR}_{1ij}\ \overline{d}_R^{i}\gamma^{\mu}e^{j}_{R}U_{1,\mu} + x^{\overline{RR}}_{1ij}\ \overline{u}_R^{i}\gamma^{\mu}\nu^{j}_{R}U_{1,\mu}$\\ 
$\widetilde{U}_{1}$     & \multicolumn{2}{c}{$\widetilde{x}^{RR}_{1ij}\ \overline{u_{R}}^{i}\gamma^{\mu}e^{j}_{R}\widetilde{U}_{1}$}           \\ 

\multirow{2}{*}{$V_{2}$}     & $-x^{RL}_{2ij}\ (\overline{d^c_{R}}^{i}\gamma^{\mu}\nu^{j}_{L}V^{1/3}_{2,\mu} - \overline{d^c_{R}}^{i}\gamma^{\mu}e^{j}_{L}V^{4/3}_{2,\mu})$            & $-x^{RL}_{2ij}\ (\overline{d^c_{R}}^{i}\gamma^{\mu}\nu^{j}_{L}V^{1/3}_{2,\mu} - \overline{d^c_{R}}^{i}\gamma^{\mu}e^{j}_{L}V^{4/3}_{2,\mu})$          \\ 
&$ + (V^{*}x_{2}^{LR})_{ij}\ \overline{u^c_L}^{i}\gamma^{\mu}e^{j}_{R}V^{1/3}_{2,\mu} - x^{{LR}}_{2ij}\ \overline{d^c_L}^{i}\gamma^{\mu}e^{j}_{R}V^{4/3}_{2,\mu}$&$ + x^{LR}_{2ij}\ \overline{u^c_L}^{i}\gamma^{\mu}e^{j}_{R}V^{1/3}_{2,\mu} - (V^{\dagger}x_{2}^{LR})_{ij}\ \overline{d^c_L}^{i}\gamma^{\mu}e^{j}_{R}V^{4/3}_{2,\mu}$\\
$\widetilde{V}_{2}$     & \multicolumn{2}{c}{$-\widetilde{x}^{RL}_{2ij}\ 
(\overline{u^c_{R}}^{i}\gamma^{\mu}e^{j}_{L}\widetilde{V}^{1/3}_{2} - 
\overline{u^c_{R}}^{i}\gamma^{\mu}\nu^{j}_{L}\widetilde{V}^{-2/3}_{2})$} \\ 
 \multirow{2}{*}{$U_{3}$}     & $-x^{LL}_{3ij}\ \overline{d_{L}}^{i}\gamma^{\mu}e^{j}_{L}U^{2/3}_{\mu} + \sqrt{2}x^{LL}_{3ij}\ \overline{d_{L}}^{i}\gamma^{\mu}\nu^{j}_{L}U^{-1/3}_{3}$             & $x_{3ij}^{LL}\ \overline{u_{L}}^{i}\gamma^{\mu}\nu^{j}_{L}U^{2/3}_{\mu}+ \sqrt{2} x^{LL}_{3ij}\ \overline{u_{L}}^{i}\gamma^{\mu}e^{j}_{L}U^{5/3}_{\mu}$         \\ 
 &  $+(Vx^{LL}_3)_{ij}\ \overline{u_{L}}^{i}\gamma^{\mu}\nu^{j}_{L}U^{2/3}_{\mu}+\sqrt{2}(Vx^{LL}_{3})_{ij}\ \overline{u_{L}}^{i}\gamma^{\mu}e^{j}_{L}U^{5/3}_{\mu}$ & $+\sqrt{2}(V^{\dagger} x_{3ij}^{LL})\ \overline{d_{L}}^{i}\gamma^{\mu}\nu^{j}_{L}U^{-1/3}_{\mu}- (V^{\dagger}x^{LL}_{3})_{ij}\  \overline{d_{L}}^{i}\gamma^{\mu}e^{j}_{L}U^{2/3}_{\mu}$ \\
 \hline
\hline
\end{tabular*}}
\end{table*}

\noindent
{\bf Scalar Lagrangian:} The kinetic Lagrangian of a generic scalar LQ $\Phi$ can be expressed as,
\begin{align}
    \mathcal{L}_{\Phi}^{kin} = \left(D_{\mu}\Phi\right)^{\dagger}\left(D^{\mu}\Phi\right) - M_{\Phi}^2 \Phi^{\dagger}\Phi,
\end{align}
where
\begin{align}\nonumber
    D_{\mu} = \partial_{\mu} - ig_{s}\frac{\lambda^{a}}{2}G^{a}_{\mu} - ig\frac{\sigma^{k}}{2}W^{k}_{\mu} - ig^{\prime}YB_{\mu}.
\end{align}
Here, $g_{s}$ is the strong coupling constant, $g$ and $g^{\prime}$ are the electroweak couplings, $Y$ is the hypercharge (expanding the covariant derivative gives the $(\gm/Z)g\Phi\Phi$ terms). The interaction Lagrangian for $\Phi$ can be written as,
\begin{align}
\label{eq:genslq}
\mathcal{L}_\Phi^{int} =&\ 
y^{L}_{\Phi, ij}\left[\bar{q}^{i,a}_R\ell^{j}_L+ \xi_\Phi \bar{q}^{\prime\ i,a}_R\nu^{j}_L\right]\Phi^a + y^{R}_{\Phi, ij}\ \bar{q}^{i,a}_L\ell^{j}_R \Phi^a + h.c.,
\end{align}
where we do not consider the diquark interaction terms. Here, $i$ and $j=\{1,\ 2,\ 3\}$ are the quark and lepton generation indices, respectively, $a$ is the colour index, $\xi_{\phi}$ is either zero or $\pm1$, and, depending on the charge of LQ, $q$ and $q^\prime$ is either a quark or a charge-conjugated quark. 
\medskip

\noindent {\bf Vector Lagrangian:} We can write the kinetic Lagrangian for a generic vector LQ $\chi$ as,
\begin{align}
    \mathcal{L}_{\chi}^{kin} = -\dfrac{1}{2}\left(D_{\mu}\chi_{\nu} - D_{\nu}\chi_{\mu}\right)^{\dagger}\left(D^{\mu}\chi^{\nu} - D^{\nu}\chi^{\mu}\right) + M_{\chi}^2 \chi^{\dagger}_{\mu}\chi^{\mu} + i g_s \left(1-\kappa\right) \chi^{\dagger}_{\mu} T^a \chi^{\nu} G^{\mu\nu a},\label{eq:vlqanokin}
\end{align}
where $\kappa$ is the additional $g \chi \chi$ coupling. The interaction term for the vLQ $\chi$ is given as, 
\begin{align}
\label{eq:genvlq}
\mathcal{L}_\chi =&\ 
x^{L}_{\chi, ij}\left[\bar{q}^{i,a}_L\gamma^\mu \ell^{j}_L+ \xi_\chi\ \bar{q}^{\prime\ i,a}_L\gamma^\mu \nu^{j}_L\right]\chi^a_\mu +
x^{R}_{\chi, ij}\ \bar{q}^{i,a}_R\gamma^\mu \ell^{j}_R~ \chi^a_\mu + \rm{h.c.},
\end{align}
where $\xi_{\chi}$ is either zero or $\pm1$.
\medskip

\noindent {\bf Up/down-aligned Yukawa interactions:} We show the LQ interactions in Table.~\ref{tab:SLQYuk}. Since one can assume the mixing among the left-handed quarks in the SM to be either in the up or down sectors, we consider two types of LQ Yukawa interactions where the LQ couples to the left-handed quarks: up-aligned, where LQ interactions are aligned with the up-type quarks (i.e., the mixing is among the down-type quarks) and down-aligned, where LQ interactions are aligned with the down-type quarks (mixing is among the up-type quarks). For clarity, we show how the Cabibbo-Kobayashi-Maskawa (CKM) matrix elements enter the Lagrangians through quark fields (in the mass basis):
\begin{align}
    &\begin{aligned}
        {d'_{L}}^{i} &= \left[V\right]_{ij}{d_{L}}^j, \\
        \overline{d'_{L}}^{i} &= \left[V^{*}\right]_{ij}\overline{d_{L}}^j, \\
        {d'^c_L}^{i} &= \left[V^{*}\right]_{ij}{d^c_{L}}^j, \\
        \overline{d'^c_L}^{i} &= \left[V\right]_{ij}\overline{d^c_L}^{j},
    \end{aligned}
    &&
    \begin{aligned}
        {u'_{L}}^{i} &= \left[V^{\dagger}\right]_{ij}u_{L}^j, \\
        \overline{u'_{L}}^{i} &= \left[V^{T}\right]_{ij}\overline{u_{L}}^j, \\
        {u'^c_{L}}^i &= \left[V^{T}\right]_{ij}{u^{c}_{L}}^j, \\
        \overline{u'^c_{L}}^{i} &= \left[V^{\dagger}\right]_{ij}\overline{u^c_{L}}^{j},
    \end{aligned}
\end{align}
where $V$ is the CKM matrix and the primed fields are in the interaction basis.

\begin{table*}
 \caption{LQ notations and Monte Carlo codes in the {\tt .fr} model files.\label{tab:namingconv}}
\centering{\small\renewcommand\baselinestretch{1.5}\selectfont
 \begin{tabular}{lcc}
 \hline
 \toprule
LQ types & FR notation  &\hspace{1cm}Monte Carlo codes\hspace{1cm} \\
\hline
 \hline
 $S_{1} (\textbf{3},\textbf{1},\frac{1}{3})$ & {\tt s101} & $4200011$ \\
 $\widetilde{S}_{1}(\overline{\textbf{3}},\textbf{1},\frac{4}{3})$ & {\tt s114} & $4200114$ \\
 $S_{3}(\overline{\textbf{3}},\textbf{3},\frac{1}{3})$ & {\tt s304, s301, s302} & $4200034$, $4200031$, $4200032$ \\
 $R_{2}(\textbf{3},\textbf{2}, \frac{7}{6})$ & {\tt r205, r202} & $4200025$, $4200022$ \\
 $\widetilde{R}_{2}(\textbf{3},\textbf{2}, \frac{1}{6})$ & {\tt r212, r211} & $4200122$, $4200121$ \\
 $\overline{S}_{1}(\textbf{3}, \textbf{1}, -\frac{2}{3})$ & {\tt s122} & $4210212$ \\\hline
 $U_{1}(\textbf{3}, \textbf{1}, \frac{2}{3})$ & {\tt u102} & $4210012$ \\
 $\widetilde{U}_1(\textbf{3}, \textbf{1}, \frac{5}{3})$ & {\tt u115} & $4210015$ \\
 $U_{3}(\textbf{3}, \textbf{3}, \frac{2}{3})$ & {\tt u305, u302, u301}  & $4210035$, $4210032$, $4210031$ \\
 $V_{2}(\overline{\textbf{3}}, \textbf{2}, \frac{5}{6})$ & {\tt v201, v204}  & $4210021$, $4210024$ \\
 $\widetilde{V}_{2}(\overline{\textbf{3}}, \textbf{2}, -\frac{1}{6})$ & {\tt v212, v211} & $4210122$, $4210121$ \\
 $\overline{U}_{1}(\textbf{3}, \textbf{1}, -\frac{1}{3})$ & {\tt u121} & $4210211$ \\
 \bottomrule
 \hline
 \end{tabular}}
\end{table*}

\subsection{FeynRules models: notations and conventions}\label{subsec:FRmodels}
\noindent
{\bf Naming convention:}
In the {\tt .fr} ({\sc FeynRules}~\cite{Alloul:2013bka}) files, the LQs are named according to the following convention (see Table~\ref{tab:namingconv}):

\begin{itemize}
    \item First character is a letter. For scalar LQs, it is either a lowercase {\tt s} (for $S_1, S_3, \widetilde{S}_1$, and $\overline{S}_{1}$) or an {\tt r} (for $R_2$ and $\widetilde{R}_1$). Similarly, for vector LQs, the letter is either a lowercase {\tt u} (for $U_1, U_3, \widetilde{U}_1$, and $\overline{U}_{1}$) or a {\tt v} (for $V_2$ and $\widetilde{V}_2$). 
    \item The next three characters are numbers. The first digit indicates whether the LQ is a singlet ($1$), doublet ($2$) or triplet ($3$) under $SU(2)_{L}$.
    \item The second digit is $1$ if there is a tilde on top of the LQ symbol, $2$ if there is a bar, and $0$ if neither.
    \item The last digit is set equal to to $\left|3Q \right|$, where $Q$ is the electric charge of the LQ.
\end{itemize}
{\bf Monte Carlo codes:} We use a similar scheme for assigning Monte Carlo codes to the LQs.

\begin{itemize}
    \item For all LQs, the first two digits are set to 42.
    \item The third digit is $0$ if the LQ is a scalar and $1$ if it is a vector.
    \item The fourth digit is kept free and set to $0$.
    \item The fifth digit is $1$ if there is a tilde on top of the LQ symbol, $2$ if there is a bar, and $0$ if neither.
    \item The sixth digit indicates the weak representation of the LQ species, i.e., it is $1$ for a singlet, $2$ for a doublet, and $3$ for a triplet.
    \item The last digit is set equal to $\left|3Q \right|$.
\end{itemize}
In the {\tt .fr} files, the particles are defined in the \texttt{M\$ClassesDescription} block. We can consider the example of $U_1$:\\
\begin{lstlisting}[frame=single,linewidth=\textwidth]
M$ClassesDescription = {
V[100] == {
    ClassName        -> u102,
    SelfConjugate    -> False,
    Indices          -> {Index[Colour]},
    Mass             -> {Mu102, 1000},
    Width            -> {Wu102, 10},
    QuantumNumbers   -> {Q -> 2/3, LeptonNumber -> -1},
    PropagatorLabel  -> "u102",
    PropagatorType   -> Sine,
    PropagatorArrow  -> Forward,
    PDG              -> 4210012,
    ParticleName     -> "u102",
    AntiParticleName -> "u102~",
    FullName         -> "up-type vector LQ"
    },
V[110] == {
    ClassName      -> u10,
    Unphysical     -> True,
    Indices        -> {Index[SU2S], Index[Colour]},
    FlavorIndex    -> SU2S,
    SelfConjugate  -> False,
    QuantumNumbers -> {Y -> 2/3},
    Definitions    -> {u10[mu_,1,cc_] :> u102[mu,cc]}
  }
};
\end{lstlisting}
The default values of the mass and decay width of \texttt{V[100]} (i.e., $U_1$) are set to $1000$ and $10$ GeV, respectively. \texttt{V[110]} is a weak-singlet (set via a user-defined index \texttt{SU2S}) unphysical field defined to include the interactions of $U_1$ with the SM gauge bosons without explicitly writing them out in the Lagrangian. 
\medskip

\noindent
{\bf LQ Yukawa couplings:} We write a generic LQ Yukawa coupling in the following form,
\begin{align*}
    y^{cd}_{ab,ij}/x^{cd}_{ab,ij}, 
\end{align*}
where
\begin{itemize}
    \item The symbol $y$ denotes a scalar LQ and $x$, a vector.
    \item The superscripts $c,d =\{L,R\}$ denote the quark and lepton chiralities, respectively.
    \item The subscript $a$ is $1$ for a weak-singlet LQ, $2$ for a doublet and $3$ for a triplet.
    \item The next subscript $b$ is $1$ if there is a tilde symbol on top of the LQ symbol, $2$ if there is a bar symbol and $0$  otherwise.
    \item The subscripts $i$ and $j$ show the quark and lepton generations, respectively.
\end{itemize}
We can consider the example of $U_{1}$, for which the Yukawa coupling matrices take the form:
\begin{align}
x^{LL}_{1} &=
\begin{bmatrix}
  x^{LL}_{10,11} & x^{LL}_{10,12} & x^{LL}_{10,13} \\
  x^{LL}_{10,21} & x^{LL}_{10,22} & x^{LL}_{10,23} \\
  x^{LL}_{10,31} & x^{LL}_{10,32} & x^{LL}_{10,33} \\
\end{bmatrix},
& 
x^{RR}_{1} &=
\begin{bmatrix}
  x^{RR}_{10,11} & x^{RR}_{10,12} & x^{RR}_{10,13} \\
  x^{RR}_{10,21} & x^{RR}_{10,22} & x^{RR}_{10,23} \\
  x^{RR}_{10,31} & x^{RR}_{10,32} & x^{RR}_{10,33} \\
\end{bmatrix}.
\label{eq:couplings}
\end{align}
The coupling, $x^{LL}_{10,21}$, couples the $U_{1}$ LQ with a second-generation quark and a first-generation lepton, and so on. In general, these Yukawa coupling matrices are complex. In the model files, the $x^{cd}_{ab,ij}/y^{cd}_{ab,ij}$ couplings are written as {\tt XABCD[I,J]/YABCD[I,J]} in the \texttt{M\$Parameters} block. For example,
\begin{lstlisting}[frame=single,linewidth=\textwidth]
M$Parameters = {
X10LL == {
    	ParameterType    -> External,
	ComplexParameter -> False,
    	Indices          -> {Index[Generation], Index[Generation]},
	BlockName        -> YUKU1LL,
    	Value            -> {X10LL[1,1] -> 0.0, X10LL[1,2] -> 0.0, X10LL[1,3] -> 0.0,
                      	     X10LL[2,1] -> 0.0, X10LL[2,2] -> 0.0, X10LL[2,3] -> 0.0,
                             X10LL[3,1] -> 0.0, X10LL[3,2] -> 0.0, X10LL[3,3] -> 0.0},
    	TeX              -> Superscript[Subscript[x,10],LL],
	InteractionOrder -> {QLD, 1},
    	Description      -> "U1 leptoquark LL Yukawa coupling matrix"
}
};
\end{lstlisting}
The model files have the following interaction hierarchy: 
\begin{lstlisting}[frame=single,linewidth=\textwidth]
M$InteractionOrderHierarchy = {
                               {QCD, 1},
                               {QED, 2},
                               {QLD, 3}
                              };
\end{lstlisting}
where {\tt QLD} is for the LQ Yukawa (i.e., new-physics) couplings. The interaction and kinetic terms are included in the Lagrangian in the following manner (considering the example of the up-aligned $U_1$ model)~\cite{Alloul:2013bka}:
\begin{lstlisting}[frame=single,linewidth=\textwidth]
Lu1Kin := Block[{mu,nu,a,aa}, 
          ExpandIndices[-(1/2)(DC[u10bar[nu,a,aa],mu] - DC[u10bar[mu,a,aa],nu]).
          (DC[u10[nu,a,aa],mu] - DC[u10[mu,a,aa],nu])]]; 

Lu1int := X10LL[i,j] u102[mu,a] (CKM[i,k].dqbar[p,k,a].Ga[mu,p,q].
          ProjM[q,r].l[r,j] + uqbar[p,i,a].Ga[mu,p,q].ProjM[q,r].vl[r,j]) + 
          X10RR[i,j] u102[mu,a] dqbar[p,i,a].Ga[mu,p,q].ProjP[q,r].l[r,j]; 

LBSM := Lu1Kin + Lu1int + HC[Lu1int];
\end{lstlisting}
The model files [in {\tt .fr} and {\sc Universal Feynman Output} (UFO)~\cite{Degrande:2011ua,Darme:2023jdn} formats] are available from the \textsc{TooLQit} repository in the directory `\texttt{FR\_models}'. 

\begin{figure}[t]
        \centering
        \includegraphics[width=0.74\linewidth]{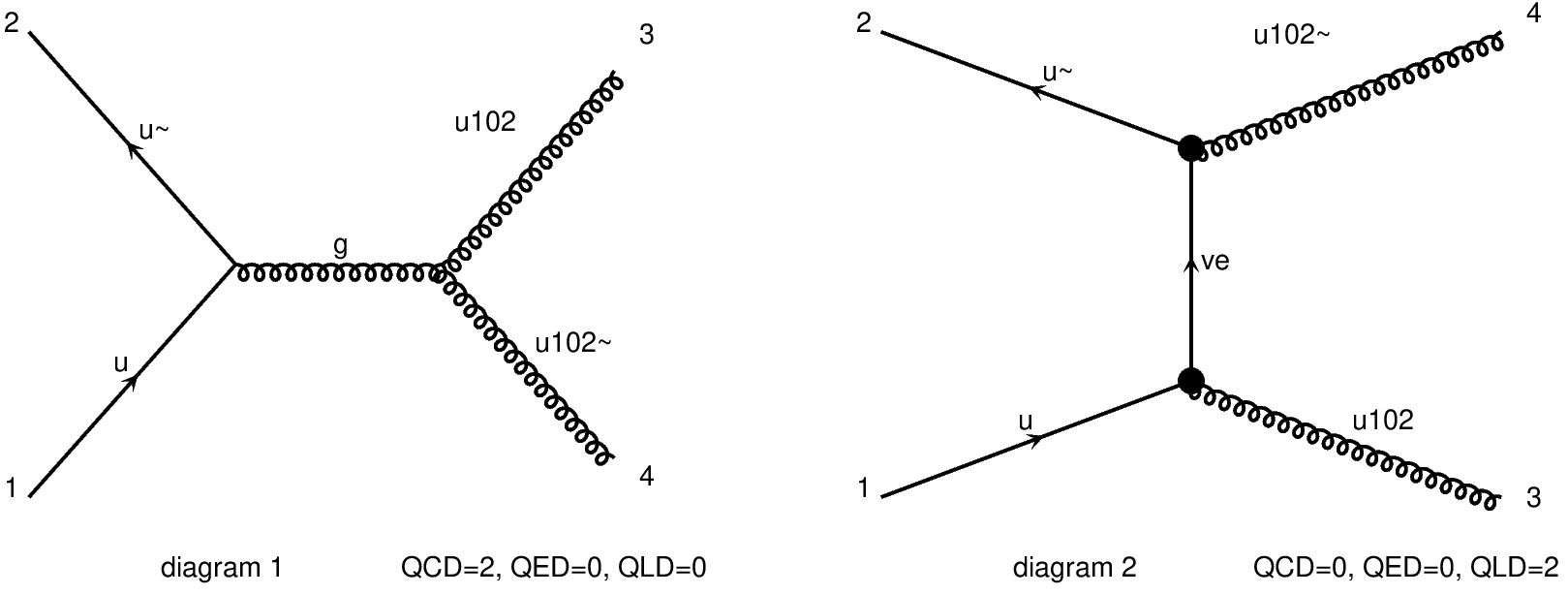}
        \label{fig:sub1}
    \caption{$U_1$ pair production at the LHC: Examples of Feynman diagrams generated by {\sc MadGraph}. 
    }
    \label{fig:pairprod}\medskip

    \centering
    \includegraphics[width=0.74\linewidth]{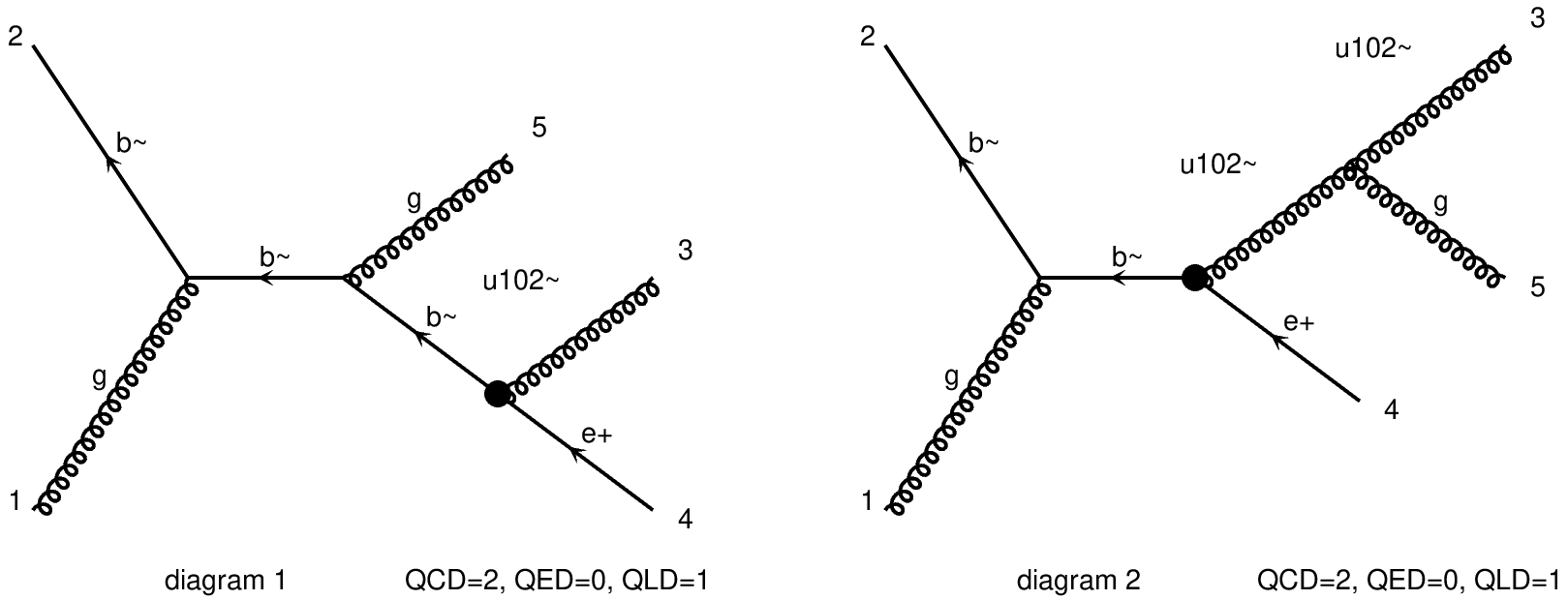}
    \caption{Three-body single production of $U_1$ at the LHC: Examples of Feynman diagrams generated by  {\sc MadGraph}. 
    }\label{fig:singleprod}\medskip

    \centering
    \includegraphics[width=0.37\textwidth]{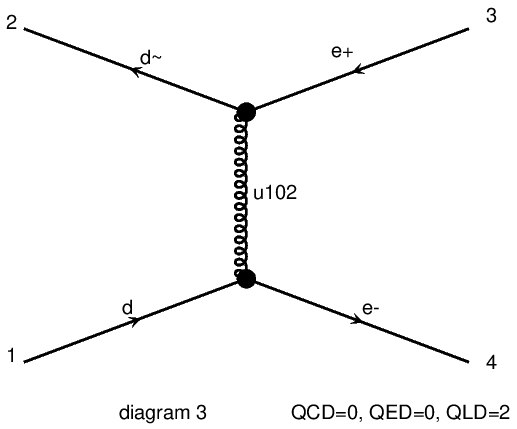}
    \caption{Drell-Yan via $U_1$: Examples of Feynman diagrams generated by  {\sc MadGraph}.
    }
    \label{fig:drellyan}
\end{figure}

\subsection{Producing $U_1$ at the LHC: Demonstration with {\sc MadGraph}}
\label{subsec:madgraph}
\noindent
For an illustration, we import the $U_1$ model UFO file~\cite{Degrande:2011ua,Darme:2023jdn} into \textsc{MadGraph5}~\cite{Alwall:2014hca} and generate the pair production process for $U_1$ at the LHC~\cite{Blumlein:1998ym} through the following command:
\begin{align}\nonumber
\texttt{generate p p > u102 u102$\sim$  QCD=2 QED=0 QLD=2}
\end{align}
which involves the QCD and LQ Yukawa couplings (see Fig.~\ref{fig:pairprod}). The $U_1$ pair can be further decayed to symmetric and asymmetric final states:
\begin{align}
    U_1 U_1 \to\left\{\begin{array}{lcl}
\multicolumn{3}{c}{\mbox{Symmetric final states}}   \\
(\ell j)(\ell j)/(\ell b)(\ell b) &\equiv& \ell \ell + 2j/2j_b \\
(j \nu)(j \nu)/(t \nu)(t \nu) &\equiv& 2j/2j_t + \slashed{E_T} \\
\multicolumn{3}{c}{\mbox{Asymmetric final states}}   \\
(\ell b) \, (\ell j) &\equiv& \ell\ell + j_b + j\\
(\ell j/\ell b) \, (j \nu/t \nu) &\equiv& \ell + (j j)/(j j_b)/(j_t j)/(j_t j_b) + \slashed{E_T}\\
(t \nu) \, (j \nu) &\equiv& j_t + j + \slashed{E_T} \\
\end{array}\right\},
\end{align}
where $\ell = e, \mu, \tau$ and $j_b,\; j_t$ denote $b$ and $t$ jets, respectively. Similarly, we can generate the $U_1$ single production processes and the $U_1$-mediated dilepton processes including its interference with the $Z/\gamma$ mediated Drell-Yan process in \textsc{MadGraph5} at the LO.

\section{C\lowercase{a}LQ: Calculator for (indirect) LHC limits}\label{sec:calq}
\noindent 
There are two main sources of LHC limits on LQ parameters~\cite{Bhaskar:2023ftn}: direct searches and the high-$P_T$ tails of the $\ell\ell$ or $\ell+\slashed{E}_T$ data. \textsc{CaLQ} is a {\sc Python} code that estimates whether the indirect LHC limits (from the high-$P_T$ tails of the dilepton data~\cite{Aad:2020zxo,Sirunyan:2021khd}) allow/exclude a point on the LQ parameter space.  It is currently at the alpha stage. While the code is generic, it supports only two LQ models---the singlet scalar $S_1$ and vector $U_1$---and has no mixed-generation dilepton or lepton plus missing energy data. In the coming versions, We plan to introduce other common LQ models and the limits from mixed generation dilepton data and direct searches (see, e.g., Refs.~\cite{Bhaskar:2021pml,Bhaskar:2023ftn}). As mentioned in the Introduction, \textsc{CaLQ} follows the $\chi^2$ minimisation and parameter limit estimation method described in Ref.~\cite{Bhaskar:2021pml} to obtain the indirect limits---it is essentially an automation of that technique. 

The $\chi^2= \chi^2(M_{LQ}, \Vec{\lm})$ function is estimated as,
    \begin{equation}
    \chi^2(M_{LQ},\Vec{\lambda}) =\sum_{\ell\ell=ee,\m\m,\ta\ta}\chi^{2}_{\ell}(M_{LQ},\Vec{\lambda}) = \sum_{\ell\ell}\left.\sum_{b\in {\rm bins}}\left(\frac{\mathcal{N}_{\rm Theory}^b(M_{LQ}, \Vec{\lm})-\mathcal{N}_{\rm Data}^b}{\Delta \mathcal{N}^b}\right)^2\right|_{\ell\ell},
    \label{eq:chisquare}    
    \end{equation}
where $\vec{\lm}=\{x_i\text{~or~}y_i\}$ denotes the set of nonzero LQ Yukawa couplings, and $\Delta \mc N = \sqrt{\left( \Delta \mc N_{stat}\right)^2 + \left( \Delta \mc N_{syst} \right)^2}$ is the error with the statistical error set as $\Delta \mc N^b_{stat} = \sqrt{\mc N_{\rm Data}^b}$ as a first approximation and an overall systematic error, i.e., $\Delta \mc N^b_{syst} = \delta \times \mc N_{\rm Data}^b$ with $\delta = 0.1$ (default value). The code uses the binned data from the HepData repository. For the $\tau\tau$ mode, it uses the transverse mass distributions from Ref.~\cite{Aad:2020zxo}. For the other leptons, it uses the dilepton invariant-mass distributions~\cite{Sirunyan:2021khd}. In the above relation, the expected number of events is estimated as
\begin{align}
    \mathcal{N}_{\rm Theory}^b (M_{LQ}, \Vec{\lm})&= \mathcal{N}_{\rm LQ}^b (M_{LQ}, \Vec{\lm})+ \mathcal{N}_{\rm SM}^b = \big[ \mathcal{N}^{pp}(M_{LQ}, \Vec{\lm}) +\mathcal{N}^{sp}(M_{LQ}, \Vec{\lm}) + \mathcal{N}^{ip} (M_{LQ}, \Vec{\lm})\big]^b + \mathcal{N}_{\rm SM}^b.\label{eq:NT}
\end{align}
Here, $\mathcal{N}^{pp}(M_{LQ}, \Vec{\lm})$, $\mathcal{N}^{sp}(M_{LQ}, \Vec{\lm})$, and $\mathcal{N}^{ip}(M_{LQ}, \Vec{\lm})$ are the numbers of events from LQ pair production (PP), single production (SP), and indirect production (IP; Drell-Yan, i.e., $qq\to\ell\ell$ via a $t$-channel LQ exchange and its interference with the SM $qq\to\ell\ell$ process) channels, respectively. For the calculator, we simulated these processes in \textsc{MadGraph5} at the LO\footnote{For $S_1$, the QCD NLO corrections are known for the pair production process~\cite{Kramer:2004df,Mandal:2015lca,Borschensky:2020hot,Borschensky:2021hbo,Borschensky:2021jyk,Borschensky:2022xsa}. To account for that, an average $k_{\rm QCD}^{\rm NLO}$ factor of $1.5$ is included for this process. This value is editable. Also, for $U_1$, we have assumed zero contribution from the additional $g\chi\chi$ coupling, $\kp$ [see Eq.~\eqref{eq:vlqanokin}].} with NNPDF2.3LO parton distributions~\cite{Ball:2012cx} and the \emph{dynamic renormalization and factorization scales} choice to estimate their contributions. The simulated events were passed through \textsc{Pythia8}~\cite{Bierlich:2022pfr} for showering and hadronisation and were matched up to two jets using the {\tt MLM} matching scheme~\cite{Mangano:2006rw,Hoche:2006ph}. Then they were passed through \textsc{Delphes}~\cite{deFavereau:2013fsa} for detector effects. We used the anti-$k_{T}$~\cite{Cacciari:2008gp} jet algorithm in \textsc{FastJet}~\cite{Cacciari:2011ma} for forming the jets. We mimicked the selection criteria and cuts used in Refs.~\cite{Aad:2020zxo,Sirunyan:2021khd} to analyse the \texttt{.root} files and obtain the resulting binwise efficiencies.  

As shown in Appendix A of Ref.~\cite{Bhaskar:2021pml}, the $\vec\lm$-dependence of the BSM contributions can be parametrised simply. For example,  $\mathcal{N}^{nr,\; b}(M_{LQ},\vec \lm)$ can be written as
\begin{align}
    \mathcal{N}^{ip,\; b}(M_{LQ}, \Vec{\lm}) = \left\{\sum_{i}^n \lm_i^2\sigma^{{ip}_2}_i(M_{LQ})\times\epsilon^{{ip}_2,\; b}_i(M_{LQ})+\sum_{i\geq j}^n \lm_i^2 \lm_j^2\sigma^{{ip}_4}_{ij}(M_{LQ})\times\epsilon^{{ip}_4,\; b}_{ij}(M_{LQ})\right\}\times \mathcal{L},
\end{align}
where $\sigma^{{ip}_4}_{ij}(M_{LQ})$ is the $t$-channel LQ exchange contribution to the dilepton cross-section calculated by setting $\lm_i=\lm_j=1$ and $\lm_{k\neq\{i,j\}}=0$, $\sigma^{{ip}_2}_{i}(M_{LQ})$ is the interference contribution obtained by setting $\lm_i=1$ and $\lm_{k\neq i}=0$, the $\epsilon$'s are the corresponding signal efficiencies (the signal fractions surviving the cuts and the detector effects in bin $b$), and $\mc L$ is the luminosity. Here, we have assumed all couplings to be real for simplicity (the LHC data is largely insensitive to the complex nature of the couplings, anyway). For a particular value of $M_{LQ}$, the $\chi^{2}$ is minimised in a $n$-dimensional space (where $n$ is the number of nonzero new Yukawa couplings). From the minimum $\chi^{2}$ value, the $1\sigma$ and $2\sigma$ parameter limits are estimated by calculating $\Delta\chi^2$. \textsc{CaLQ} uses interpolated cross-sections and efficiencies from stored data files.

\subsection{Setting up the calculator}
\noindent To use \textsc{CaLQ}, one can clone the \textsc{TooLQit} repository and access the \textsc{CaLQ} directory using the following commands:

\begin{lstlisting}[frame=single,linewidth=0.8\textwidth]
$ cd <folder_to_clone_TooLQit>
$ git clone https://github.com/rsrchtsm/TooLQit.git
$ cd TooLQit/CaLQ/Version_X.Y.Z
\end{lstlisting} 

\noindent Or, one can also download the zip file from \href{https://github.com/rsrchtsm/TooLQit/archive/refs/heads/main.zip}{https://github.com/rsrchtsm/TooLQit/archive/refs/heads/main.zip} and unzip the file.

\medskip

\noindent \textsc{CaLQ} is a {\sc Python3} code. It is possible to use it in a virtual environment or directly. It depends on four core packages (\textsc{numpy}, {\sc scipy}, {\sc sympy}, and {\sc pandas}) and the \textsc{prompt\_toolkit} package for auto-completion on the command line. The command
\begin{lstlisting}[frame=single,linewidth=0.8\textwidth]
$ pip install numpy sympy scipy pandas prompt_toolkit
\end{lstlisting}
installs the required packages without a virtual environment.
Otherwise, we can create a virtual environment:
\begin{lstlisting}[frame=single,linewidth=0.8\textwidth]
$ python3 -m venv venv
$ source venv/bin/activate
$ pip install -r requirements.txt
\end{lstlisting}
\textsc{CaLQ} is now ready for use. 

\subsection{Running the calculator}
\noindent There are two ways to use \textsc{CaLQ}: interactive and non-interactive. 
\medskip

\noindent 
{\bf Interactive mode:}
The interactive mode is useful for testing a few parameter points. Entering the following command takes us to the interactive mode. 
\begin{lstlisting}[frame=single,linewidth=0.8\textwidth]
$ python3 calq.py
\end{lstlisting}
The \textsc{CaLQ} logo appears (see Fig.~\ref{fig:interactive_1}). It is followed by a list of available input commands, the supported LQ models and some illustrative Yukawa couplings to show the format. 
\begin{itemize}
\item \lstinline{Couplings available}: The format of the coupling(s) are mentioned in Section.~\ref{subsec:FRmodels}. 
The calculator-specific $U_1$ and $S_1$ couplings are listed below as matrices:
\begin{align}
x^{LL}_{1} =&
\begin{bmatrix}
  \texttt{X10LL[1,1]} & \texttt{X10LL[1,2]} & \texttt{X10LL[1,3]} \\
  \texttt{X10LL[2,1]} & \texttt{X10LL[2,2]} & \texttt{X10LL[2,3]} \\
  \texttt{X10LL[3,1]} & \texttt{X10LL[3,2]} & \texttt{X10LL[3,3]} \\
\end{bmatrix},\label{eq:leftcouplingsX}\\
x^{RR}_{1} =&
\begin{bmatrix}
  \texttt{X10RR[1,1]} & \texttt{X10RR[1,2]} & \texttt{X10RR[1,3]} \\
  \texttt{X10RR[2,1]} & \texttt{X10RR[2,2]} & \texttt{X10RR[2,3]} \\
  \texttt{X10RR[3,1]} & \texttt{X10RR[3,2]} & \texttt{X10RR[3,3]} \\
\end{bmatrix},\label{eq:rightcouplingsX}\\
y^{LL}_{1} =&
\begin{bmatrix}
  \texttt{Y10LL[1,1]} & \texttt{Y10LL[1,2]} & \texttt{Y10LL[1,3]} \\
  \texttt{Y10LL[2,1]} & \texttt{Y10LL[2,2]} & \texttt{Y10LL[2,3]} \\
  \texttt{-} & \texttt{-} & \texttt{-} \\
\end{bmatrix},\\
y^{RR}_{1} =&
\begin{bmatrix}
  \texttt{Y10RR[1,1]} & \texttt{Y10RR[1,2]} & \texttt{Y10RR[1,3]} \\
  \texttt{Y10RR[2,1]} & \texttt{Y10RR[2,2]} & \texttt{Y10RR[2,3]} \\
  \texttt{-} & \texttt{-} & \texttt{-} \\
\end{bmatrix}.\label{eq:rightcouplingsY}
\end{align}
The $S_1$ couples to a charged lepton along with an up-type quark. The current \textsc{CaLQ} does not put limits on the top-quark couplings (\texttt{Y10XX[3,J]}) as the top quark is essentially absent in the initial states.
\begin{figure}[t]
    \centering
    \includegraphics[width=\textwidth]{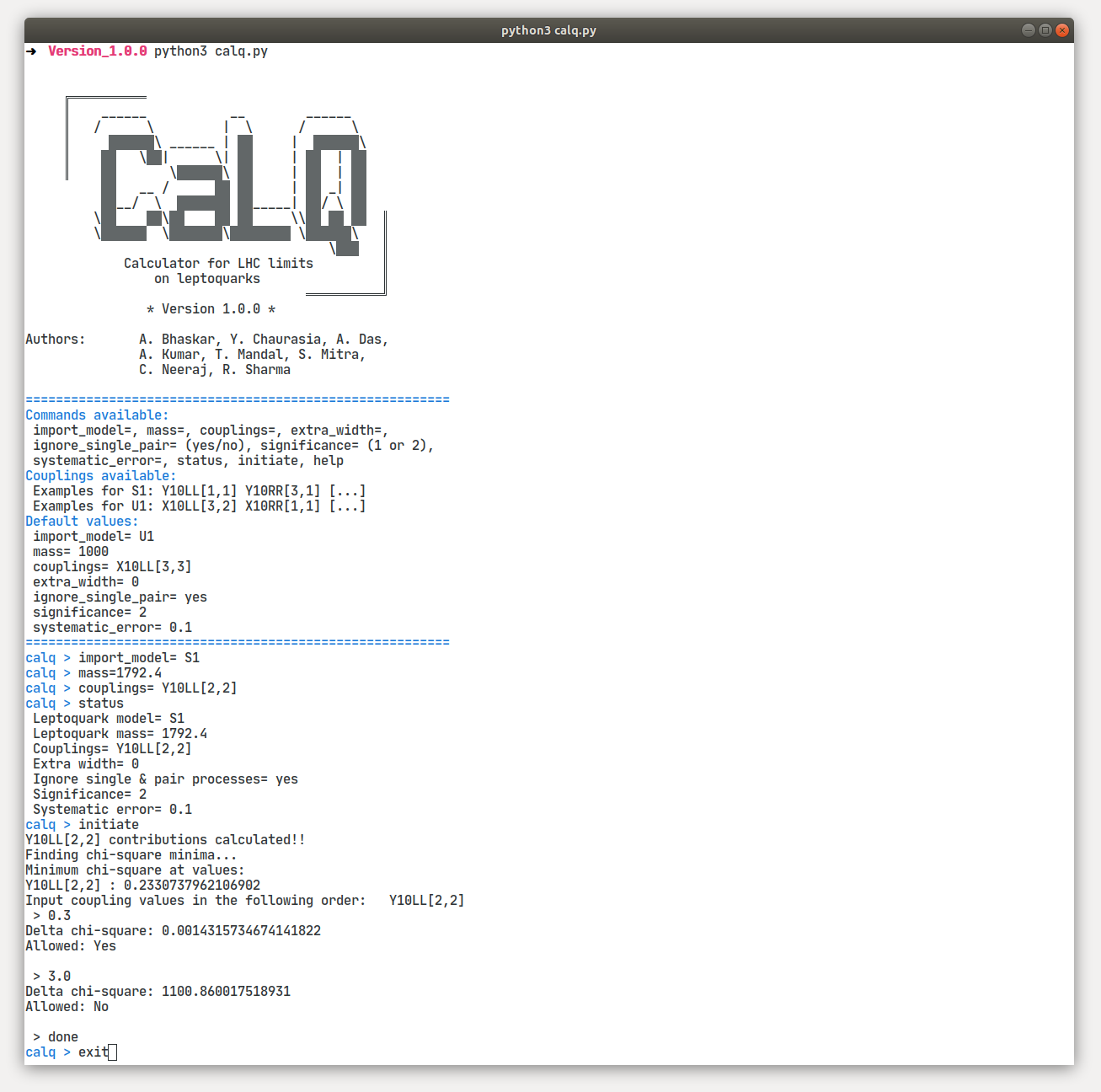}
    \caption{Screenshot of \textsc{CaLQ} running in the interactive mode.}
    \label{fig:interactive_1}
\end{figure}
\item \lstinline{ignore_single_pair}: This command allows the user to ignore the resonant pair and single production contributions. For heavy LQs, the contribution from the resonant modes is small. The choices are `\texttt{yes}'/`\texttt{no}' or `\texttt{y}'/`\texttt{n}'. Inputting `\texttt{yes}' tells  \textsc{CaLQ} to ignore the resonant contribution to evaluate the limits; this helps in speeding up the calculations. The default input is set to `\texttt{yes}'.

\item \lstinline{significance}: Input $1$ and $2$, for $1\sigma$ and $2\sigma$ limits, respectively. 

\item \lstinline{systematic_error}: The fractional systematic error, $\delta = \delta^b$ appearing in $\Delta \mc N^b_{syst}$ [Eq.~\eqref{eq:chisquare}]. The default value is $0.1$.

\item \lstinline{extra_width}: The indirect limits are largely independent of branching ratios (BRs). However, for a light LQ, the single and pair production processes contribute to the dilepton final states, which depend on the BRs. \textsc{CaLQ} estimates the relevant BRs automatically from the choice of Yukawa couplings and a hardcoded (approximate, LO) decay width expression. If, however, there is an additional decay mode, the user can add the extra width in GeV. The default value is $0$ GeV.

\end{itemize}
The list is followed by a prompt, `\texttt{calq >}'. Then the following inputs initialise the calculator.
\begin{itemize}
\item `\texttt{calq > import\_model=}': The choice of LQ, `\texttt{S1}' or `\texttt{U1}'. 

\item `\texttt{calq > mass=}': The mass of the LQ in GeV. Currently, the calculator computes the LHC bounds for LQs in the mass range $1000$--$5000$ GeV. 

\item `\texttt{calq > couplings=}': The nonzero Yukawa couplings, each separated from the previous one by a space.

\item `\texttt{calq > initiate}': Initiates the calculator and computes the $\chi^2$ and its minimum(minima) corresponding to the input values and coupling(s). 
\end{itemize}
For instance, the inputs below will select a $1000$ GeV $U_1$ in a two-coupling scenario.
\vspace{0.2cm}
\begin{lstlisting}[frame=single,linewidth=0.8\textwidth]
calq > import_model= U1
calq > mass= 1000.0
calq > couplings= X10LL[1,1] X10LL[3,2]
calq > initiate
\end{lstlisting}
Once initiated, the \textsc{CaLQ} prompt changes from `\texttt{calq > }' to `\lstinline{> }'. We can now enter the values of the Yukawa couplings to be tested. The prompt accepts inputs in the form `\lstinline{<f1> <f2> }$\cdots$\lstinline{ <fn>}' (\lstinline{<f1>} to \lstinline{<fn>} are floating point numbers, i.e., real and \emph{n} is the number of Yukawa couplings, see Fig.~\ref{fig:interactive_1}). The couplings can be entered in the manner shown below:
\begin{lstlisting}[frame=single,linewidth=0.8\textwidth]
 > 0.1 0
 > 0.37 0.0001
 > 0.5 0.7
\end{lstlisting}
If there are multiple couplings, we can enter the couplings separated by a space. We enter the values of the couplings in the same order as the input couplings. Based on the $\delta\chi^2$, the allowed (disallowed) input Yukawa couplings within the $1\sigma$ or $2\sigma$ exclusion limits are displayed with a yes or no. Entering
\begin{lstlisting}[frame=single,linewidth=0.8\textwidth]
 > done
\end{lstlisting}
(or `\texttt{d}', `\texttt{q}', `\texttt{quit}', `\texttt{exit}')
exits the query mode. The prompt then returns to the input mode, and input parameters show the previous values, which can be updated.

\textsc{CaLQ} also supports the following two commands:
\begin{itemize}
\item `\lstinline{status}': The user can see the current values entered as inputs.

\item `\lstinline{help}': Displays the list of commands available.
\end{itemize}
Finally, the command
\begin{lstlisting}[frame=single,linewidth=0.8\textwidth]
calq > exit
\end{lstlisting}
(or `\texttt{q}', `\texttt{quit}', `\texttt{e}', `\texttt{.exit}', `\texttt{exit()}')
stops the calculator. \bigskip

\begin{figure*}
 \centering
    \includegraphics[width=\textwidth]{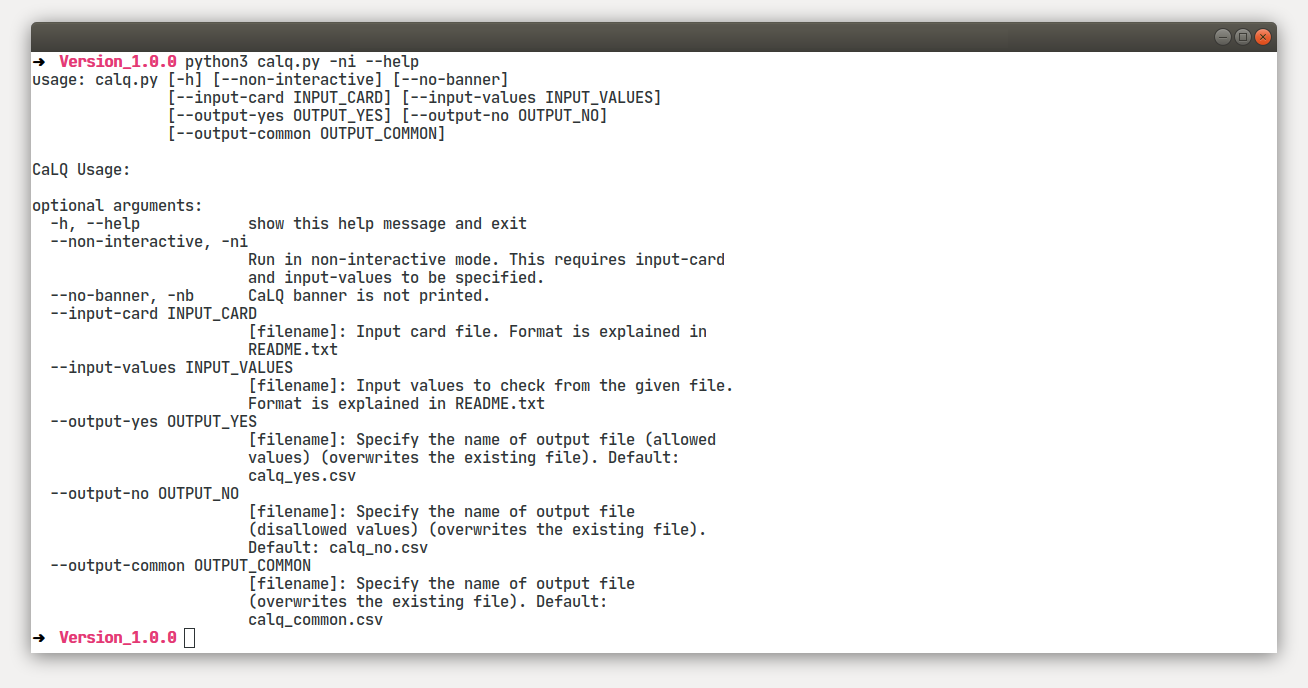}
    \caption{The \textsc{CaLQ} help menu in the non-interactive mode.\label{fig:help}}
\end{figure*}

\begin{figure*}
    \includegraphics[width=\textwidth]{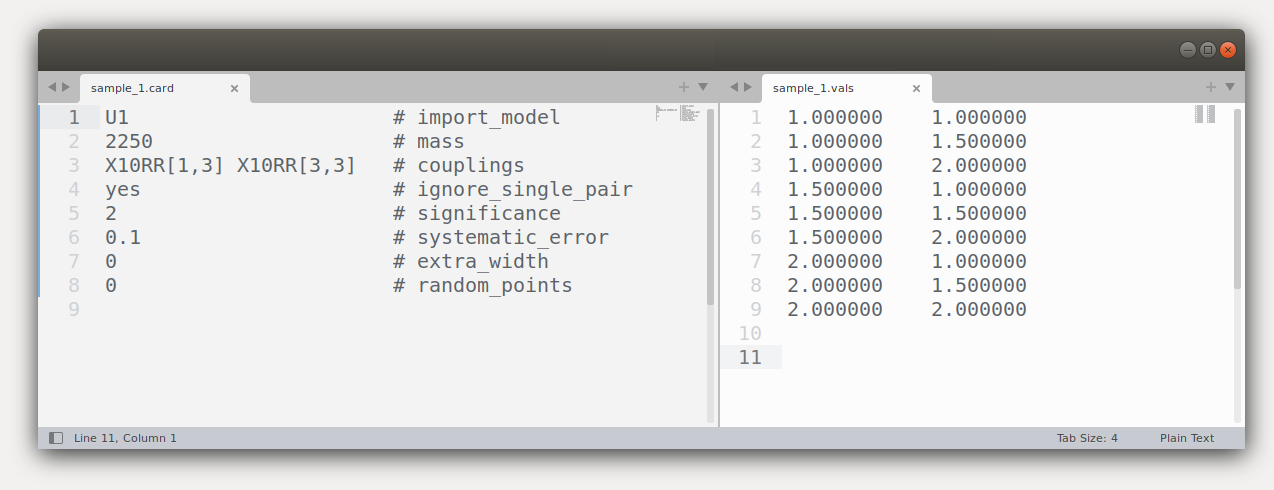}
    \caption{Non-interactive inputs.\label{fig:non-int-calQ}}

~\\

\includegraphics[width=\textwidth]{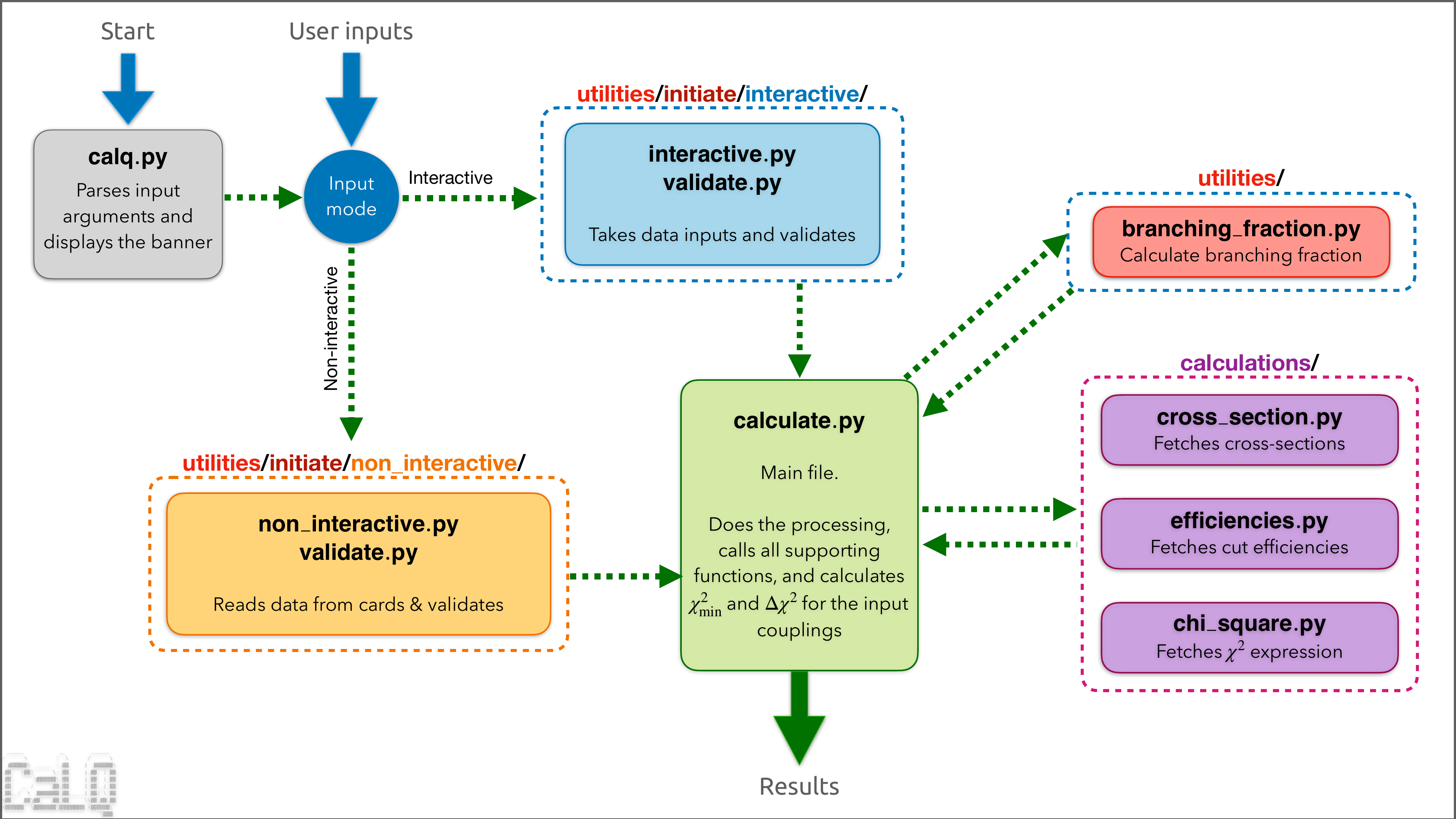}
\caption{The \textsc{CaLQ} codeflow.\label{fig:codeflow}}
\end{figure*}

\noindent{\bf Non-interactive mode}: 
To use the calculator in the non-interactive mode, we use the tag \lstinline{-ni} or \lstinline{--non-interactive}: 
\begin{lstlisting}[frame=single,linewidth=0.8\textwidth]
$ python3 calq.py -ni [options]
\end{lstlisting}
The \lstinline{options} field takes the following inputs:
\begin{itemize}
\item \lstinline{--help}: Displays the help message (see Fig.~\ref{fig:help}).
\item \lstinline{--input-card=[filename]}: Takes a file with \texttt{.card} extension where we can specify the input parameters as follows:\\~\\ 
Line 1: Model name (e.g., \texttt{S1} or \texttt{U1})\\
Line 2: LQ mass in GeV\\
Line 3: Yukawa couplings (e.g., \texttt{X10LL[1,2]} \texttt{X10LL[2,2]})\\
Line 4: \texttt{ignore\_single\_pair} (\texttt{yes} or \texttt{no})\\
Line 5: \texttt{significance} (1$\sigma$ or 2$\sigma$)\\
Line 6: \texttt{systematic\_error}\\
Line 7: \texttt{extra\_width}\\
Line 8: \texttt{random\_points} [If \texttt{random\_points} is set to zero, the user has to enter the Yukawa coupling values in a separate text file with the extension \lstinline{--input-values=[filename]}. If one set \texttt{random points} as, say, "$1000$", the calculator generates $1000$ random points between $-3.5$ and $3.5$ as inputs to the Yukawa couplings. An example input card and an example \texttt{.vals} file are found in the $\texttt{sample}$ directory, see Fig.~\ref{fig:non-int-calQ}]

\item \lstinline{--no-banner} or \lstinline{-nb}: The \textsc{CaLQ} banner is not printed.
\item \lstinline{--output-yes=[filename]}: We specify the name of the output file containing allowed parameter points (overwrites any existing file). The default name of this output file is set as  \lstinline{calq_yes.csv}. Otherwise, in the field \lstinline{filename} we can specify the path of the output file with a different name.
\item \lstinline{--output-no=[filename]}: Specify the name of the output file containing the disallowed parameter points (overwrites the existing file). The default name of this output file is set as  \lstinline{calq_no.csv}. Otherwise, in the field \lstinline{filename} we can specify the path of the output file with a different name.
\end{itemize}
A sample bash script (\texttt{sample\_1.sh}) is available in the \texttt{CaLQ/Version\_X.Y.Z} folder. After suitably modifying the input parameters in the \texttt{.card} file, one can enter the desired couplings in the \texttt{.vals} file and run the bash file to obtain the output. The non-interactive mode relies on the input card and the query values. The output files are generated in the comma-separated-values (\lstinline{.csv}) format in the order of the given couplings. The last number in a row shows the $\Delta \chi^2$ value for the particular parameter set. The \texttt{calq\_yes.csv} and \texttt{calq\_no.csv} files can be used for further analysis. 

\begin{figure*}[]
\centering
\captionsetup[subfigure]{labelformat=empty}
\subfloat{\includegraphics[width=0.23\linewidth]{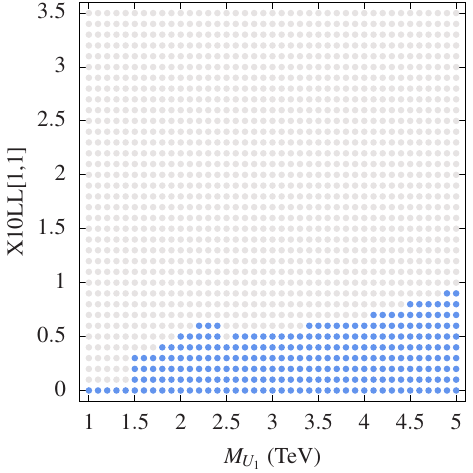}\label{fig:X10LL11}}\hspace{1cm}
\subfloat{\includegraphics[width=0.23\linewidth]{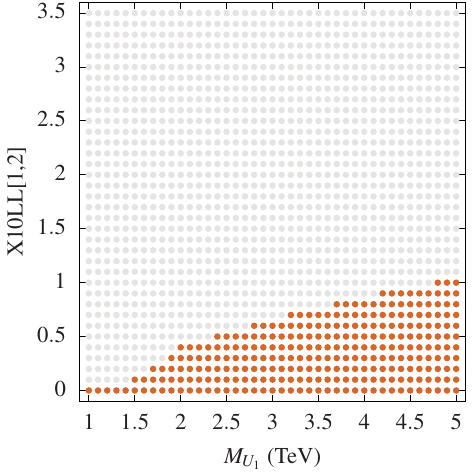}\label{fig:X10LL12}}\hspace{1cm}
\subfloat{\includegraphics[width=0.23\linewidth]{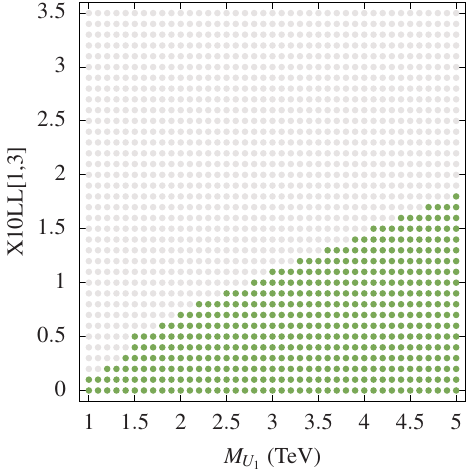}\label{fig:X10LL13}}\\
\subfloat{\includegraphics[width=0.23\linewidth]{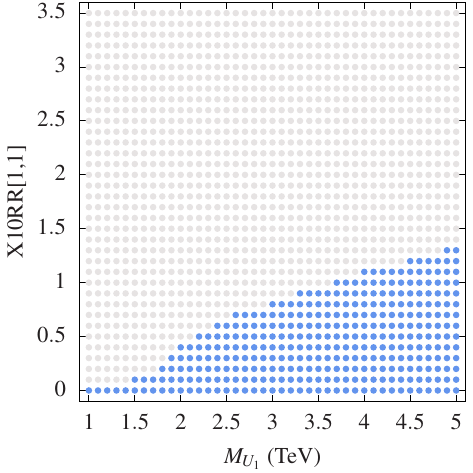}\label{fig:X10RR11}}\hspace{1cm}
\subfloat{\includegraphics[width=0.23\linewidth]{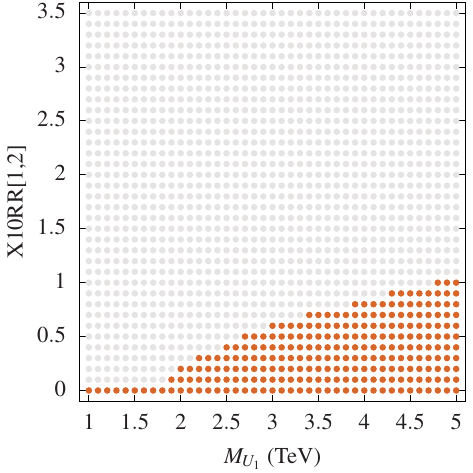}\label{fig:X10RR12}}\hspace{1cm}
\subfloat{\includegraphics[width=0.23\linewidth]{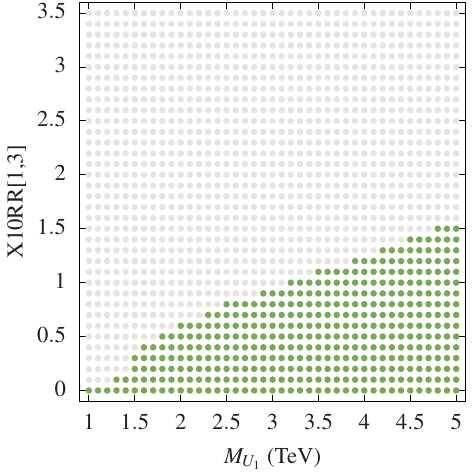}\label{fig:X10RR13}}\\
\subfloat{\includegraphics[width=0.23\linewidth]{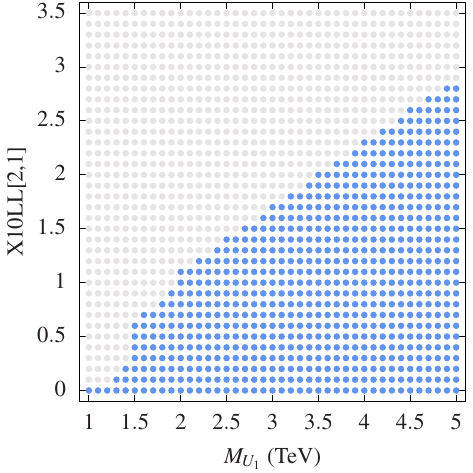}\label{fig:X10LL21}}\hspace{1cm}
\subfloat{\includegraphics[width=0.23\linewidth]{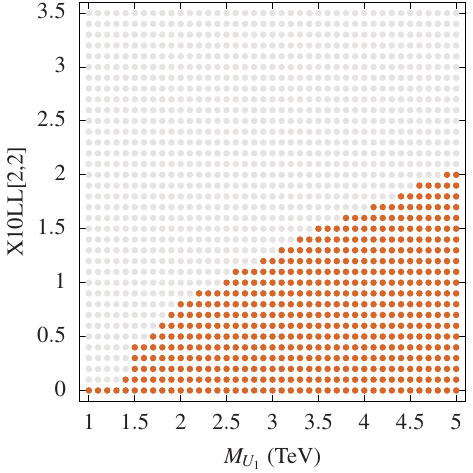}\label{fig:X10LL22}}\hspace{1cm}
\subfloat{\includegraphics[width=0.23\linewidth]{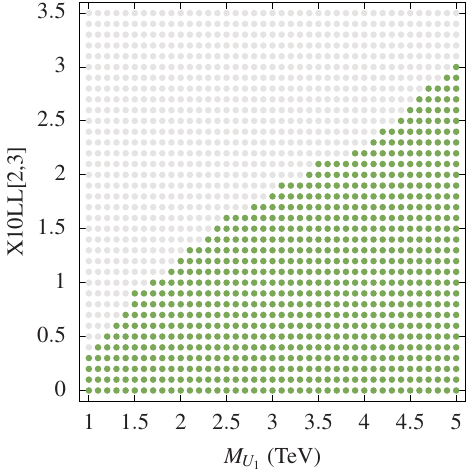}\label{fig:X10LL23}}\\
\subfloat{\includegraphics[width=0.23\linewidth]{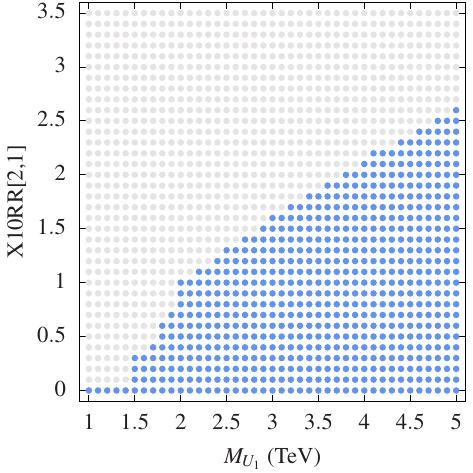}\label{fig:X10RR21}}\hspace{1cm}
\subfloat{\includegraphics[width=0.23\linewidth]{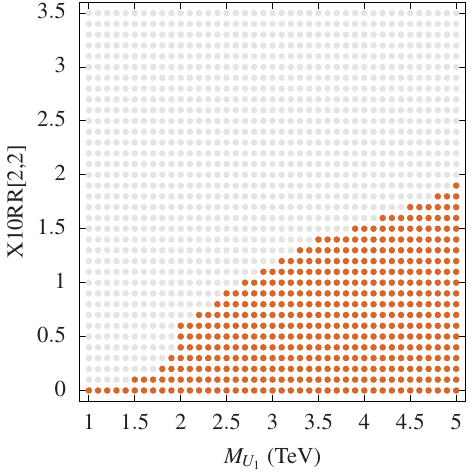}\label{fig:X10RR22}}\hspace{1cm}
\subfloat{\includegraphics[width=0.23\linewidth]{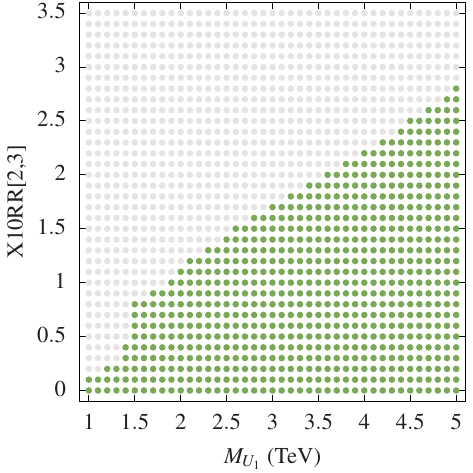}\label{fig:X10RR23}}\\
\subfloat{\includegraphics[width=0.23\linewidth]{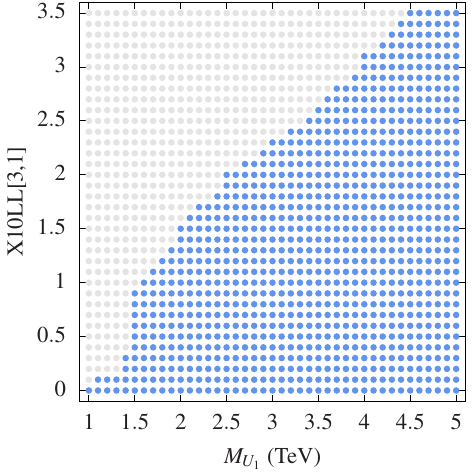}\label{fig:X10LL31}}\hspace{1cm}
\subfloat{\includegraphics[width=0.23\linewidth]{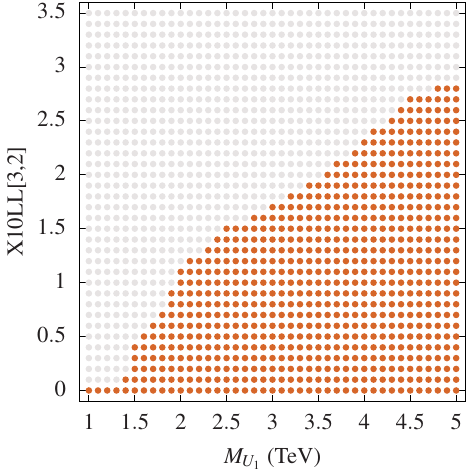}\label{fig:X10LL32}}\hspace{1cm}
\subfloat{\includegraphics[width=0.23\linewidth]{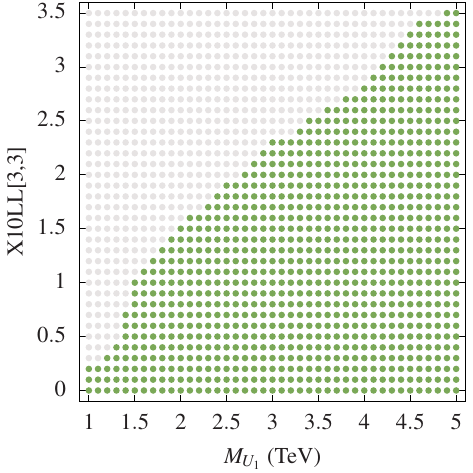}\label{fig:X10LL33}}\\
\subfloat{\includegraphics[width=0.23\linewidth]{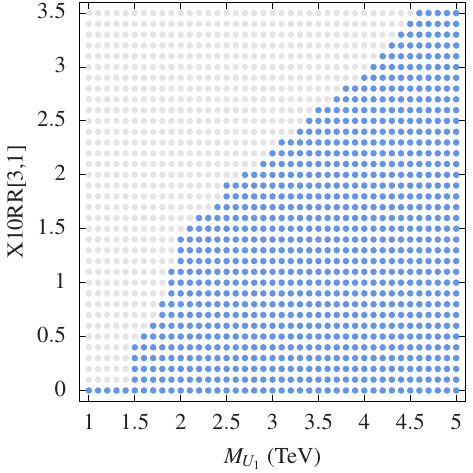}\label{fig:X10RR31}}\hspace{1cm}
\subfloat{\includegraphics[width=0.23\linewidth]{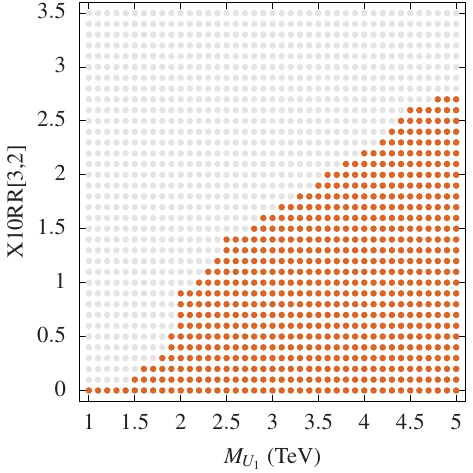}\label{fig:X10RR32}}\hspace{1cm}
\subfloat{\includegraphics[width=0.23\linewidth]{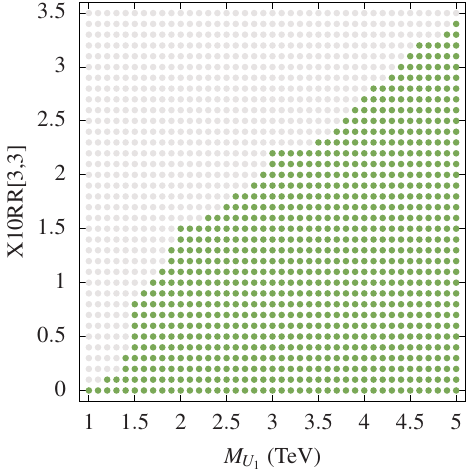}\label{fig:X10RR33}}
\caption{Illustrative one-coupling scans for the $U_1$. The grey regions are ruled out at the $2\sigma$ level.}
\label{fig:onecouplingscans}
\end{figure*}
\begin{figure*}[]
\centering
\captionsetup[subfigure]{labelformat=empty}
\subfloat{\includegraphics[width=0.23\linewidth,height=0.23\linewidth]{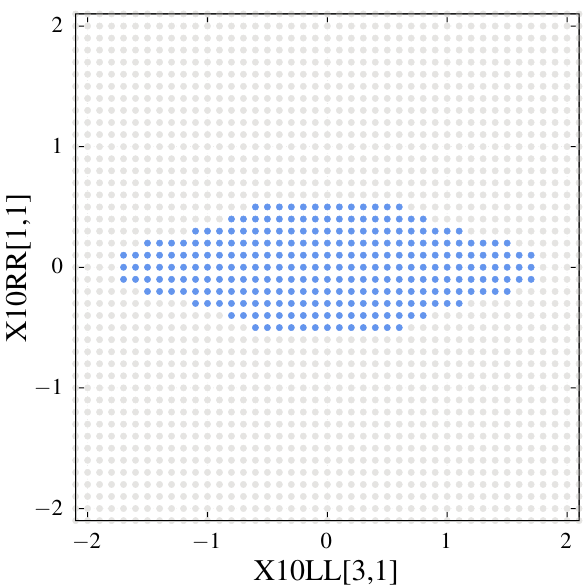}\label{fig:X10LL31_X10RR11}}\hspace{1cm}
\subfloat{\includegraphics[width=0.23\linewidth,height=0.23\linewidth]{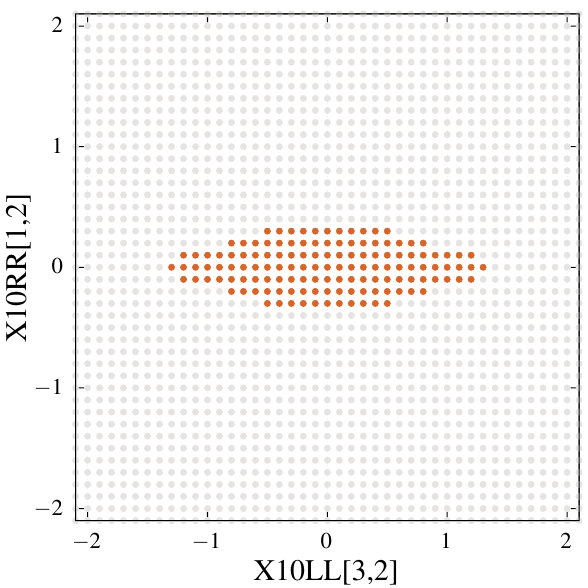}\label{fig:X10LL32_X10RR12}}\hspace{1cm}
\subfloat{\includegraphics[width=0.23\linewidth,height=0.23\linewidth]{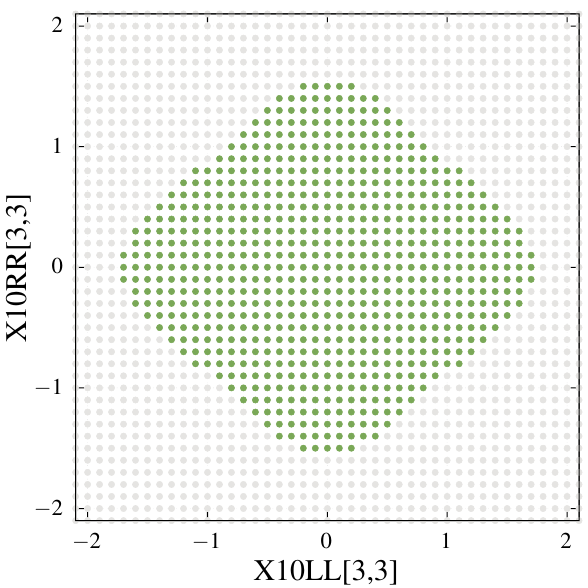}\label{fig:X10LL33_X10RR33}}\\
\subfloat{\includegraphics[width=0.23\linewidth,height=0.23\linewidth]{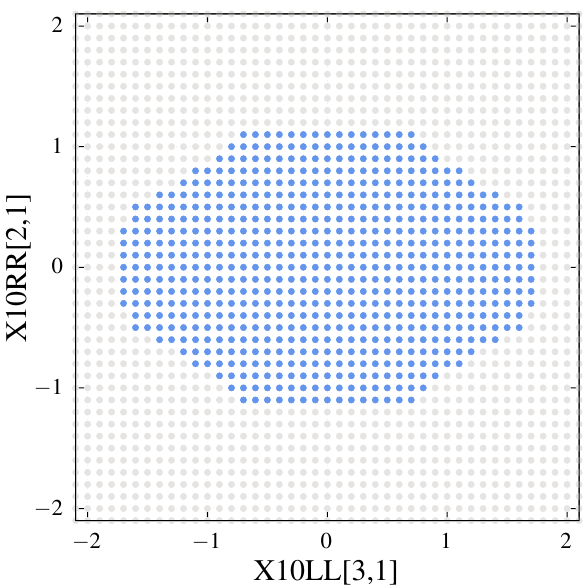}\label{fig:X10LL31_X10RR21}}\hspace{1cm}
\subfloat{\includegraphics[width=0.23\linewidth,height=0.23\linewidth]{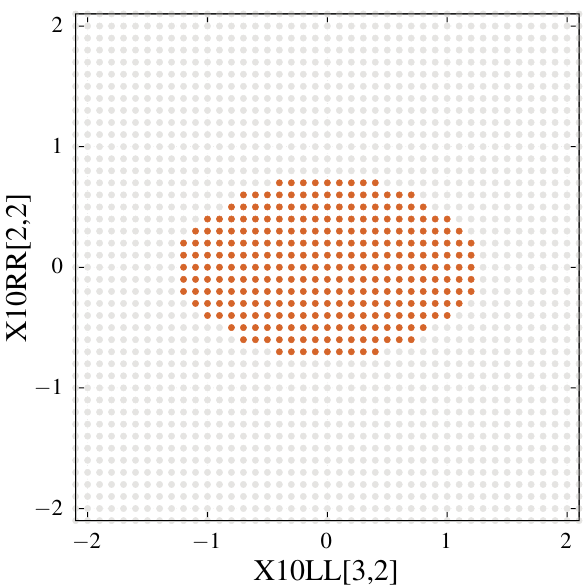}\label{fig:X10LL32_X10RR22}}\hspace{1cm}
\subfloat{\includegraphics[width=0.23\linewidth,height=0.23\linewidth]{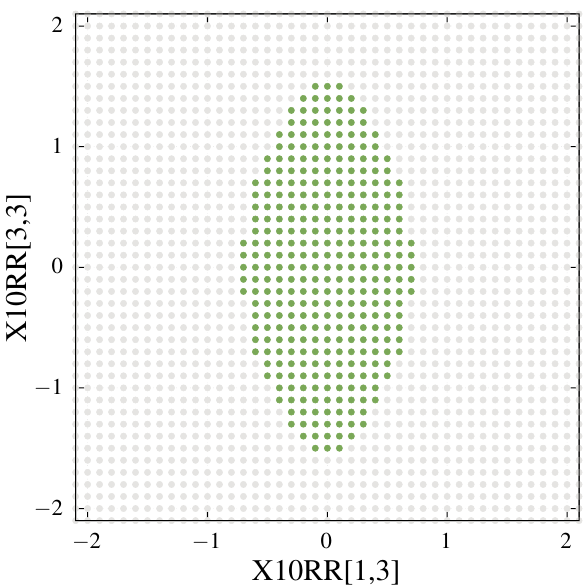}\label{fig:X10RR13_X10RR33}}\\
\subfloat{\includegraphics[width=0.23\linewidth,height=0.23\linewidth]{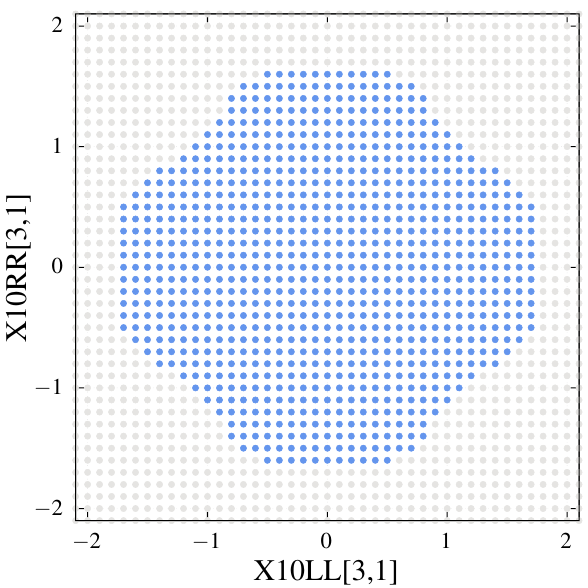}\label{fig:X10LL31_X10RR31}}\hspace{1cm}
\subfloat{\includegraphics[width=0.23\linewidth,height=0.23\linewidth]{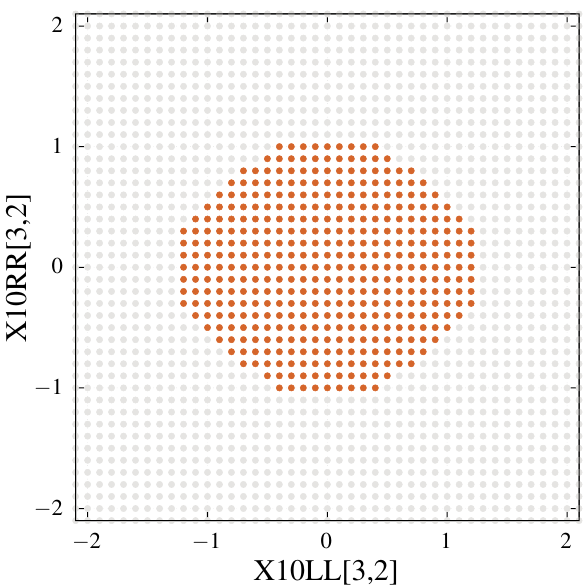}\label{fig:X10LL32_X10RR32}}\hspace{1cm}
\subfloat{\includegraphics[width=0.23\linewidth,height=0.23\linewidth]{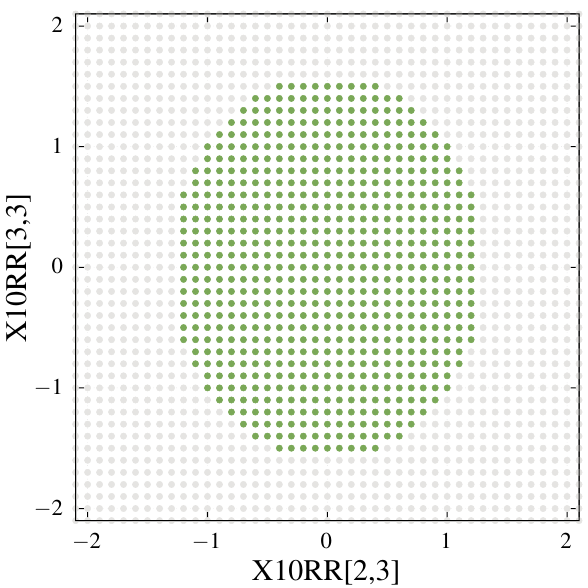}\label{fig:X10RR23_X10RR33}}
\caption{Illustrative two-coupling scans for a $2250$ GeV $U_1$. The grey regions are ruled out at the $2\sigma$ level.}
\label{fig:twocouplingscans}
\end{figure*}

\subsection{\textsc{CaLQ} workflow}
\noindent 
Once the input commands, such as the LQ model, couplings, mass, etc., are entered, fields are type-checked, validated and confirmed to be within acceptable ranges (e.g., the mass of the LQ should be within $1-5$ TeV); incorrect inputs/formats lead to error messages. From the input coupling string, \textsc{CaLQ} reads the chirality and the generation information. For instance, from the input `\texttt{couplings=X10LL[1,2]}' for the $U_1$ LQ, \textsc{CaLQ} reads the quark information (first-generation, left-handed) and the lepton information (second-generation, left-handed). Then, depending on the mass input, it fetches the relevant cross-sections of various production modes and the corresponding binwise efficiencies. If we set `\texttt{mass=2000}' (GeV) in the $U_1$ LQ example, \textsc{CaLQ} accesses the cross-section and the binwise efficiencies of the non-resonant production ($d\bar d\to \mu^-\mu^+$) for a $2000$ GeV $U_1$ LQ (\textsc{CaLQ} ignores the resonant productions to save computation by default -- those are important mainly in the low mass regions. The user has the option). The cross-section and the binwise efficiencies of the LQs are stored in steps of $500$ GeV. For intermediate values, cross-sections and efficiencies are calculated via interpolation. 

With the cross-sections and efficiencies, \textsc{CaLQ} forms the $\chi^2$ polynomial and varies the coupling(s) between $[-3.5,3.5]$ to evaluate it on the $ee$, $\mu\mu$, and $\tau\tau$ datasets and combines the results. Then, it looks for the global minimum using the \texttt{scipy.optimize()} function from \textsc{Python}’s \texttt{scipy} library. The function is called with multiple starting points to prevent it from running into a local minimum. Once the minimum  $\chi^2$ is calculated, \textsc{CaLQ} calculates the $\chi^2$ for the input couplings and, based on the number of input couplings, estimates the corresponding $\Delta\chi^2 = \chi^2 - \chi^2_{\text{min}}$ to check whether the parameter point is within the allowed range ($1\sigma$ or $2\sigma$). For example, for a single coupling, the program will output `\lstinline{yes}' if $\Delta \chi^2 = 4.0$, indicating the coupling is allowed within the $2\sigma$ range. If the coupling does not satisfy this criterion, the program will output `\lstinline{no}'. Fig.~\ref{fig:codeflow} illustrates the codeflow of \textsc{CaLQ} in detail.

\subsection{Demonstration: Limits on $U_1$ parameters}
\noindent To demonstrate the outputs of \textsc{CaLQ}, we show the results of one-coupling scenarios with $U_1$ in Fig.~\ref{fig:onecouplingscans}. For these, we passed a coupling-mass grid (with step size \{$\Delta\lm, \Delta M_{U_1}\}=\{0.1,\ 100$ GeV\}) to \textsc{CaLQ} for each coupling and marked the allowed/not allowed points with different colours. In Fig.~\ref{fig:twocouplingscans}, we show some two-coupling scans. For these, we set the mass of $U_1$ at a random value, $2250$ GeV, and perform a two-coupling grid scan.

\section{Conclusion: summary and outlooks}
\label{sec:conclu}
\noindent
We introduced the LQ toolkit, \textsc{TooLQit}, which includes LO \textsc{FeynRules} models of all possible LQs and \textsc{CaLQ}, a calculator designed to estimate indirect limits from dilepton data. This comprehensive set of models and the accompanying calculator offer valuable resources for BSM phenomenology studies and future experimental searches at the LHC. LQs, being integral components of a wide range of BSM scenarios, are actively searched for in LHC experiments. \textsc{TooLQit} represents a foundational step towards consolidating various LQ-related computational tools onto a unified platform. This toolkit lets users evaluate constraints on LQ models and explore their discovery potential at the LHC or other collider experiments. While the current version has certain limitations (detailed below), the fully open-source nature of the code provides users with flexibility and insights, allowing them to adapt and extend the tools for new/custom cases.
\medskip

\noindent 
{\bf \textsc{TooLQit}/FeynRules models:}
    \begin{itemize}
        \item[--] The set contains all LQs listed in Refs.~\cite{Buchmuller:1986zs,Dorsner:2016wpm}. The models follow a set of systematic and easy-to-follow notations/naming conventions (explained in Section~\ref{subsec:FRmodels}). Apart from the \texttt{.fr} files and the \textsc{Mathematica} codes, we also provide the \textsc{Universal FeynRules Output} files suitable for \textsc{MadGraph5}.
        
        \item[--] Currently, the models provided are at LO. While NLO QCD LQ models are already available in the literature (e.g., see Refs.~\cite{Mandal:2015lca, Korajac:2023xtv}), the \textsc{TooLQit} LO models include interactions of LQs with electroweak gauge bosons, such as the mixed QCD-QED $\gamma/Z$-$g$-LQ-LQ vertex. These interactions can be significant in some scenarios (e.g., see Ref.\cite{Bhaskar:2023ftn}), especially when the EM charge of LQ is high.
    \end{itemize} 

\noindent 
{\bf \textsc{TooLQit/CaLQ}:}    
\begin{itemize}
        \item[--] It is a {\sc Python} package that automatically estimates the indirect LHC limits on the LQ-$q$-$\ell$ Yukawa couplings by a $\chi^2$ estimation. It is based on the method we developed in Ref.~\cite{Mandal:2018kau} (applied in Ref.~\cite{Aydemir:2019ynb}) and generalised in Ref.~\cite{Bhaskar:2021pml}. 
        
        \item[--] Currently, it is at the alpha stage: it contains the data for just the two LQ models. For two weak-singlet LQs (the charge-$1/3$ scalar $S_1$ and the charge-$2/3$ vector $U_1$), CaLQ can check whether a parameter point (i.e., the mass of the LQ---between $1$ and $5$ TeV---and a set of nonzero LQ-$q$-$\ell$ couplings) is allowed by the current dilepton ($ee$, $\m\m$, $\ta\ta$) data~\cite{Aad:2020zxo,Sirunyan:2021khd}. 

        \item[--] It has a command-line interface and works in two modes: interactive and non-interactive. The interactive mode is suitable for testing a few coupling points at specific mass values. The non-interactive mode is designed to handle a (large) list of parameters, enabling users to run a scan or check whether a given LQ parameter region satisfies experimental constraints. The non-interactive mode is handy for evaluating whether a parameter space allowed by other experimental bounds is consistent with the LHC data.        
    \end{itemize} 

The $B$-meson anomalies---which have brought much attention to LQs in the recent literature---might have largely subsided in the latest measurements. However, LQs remain among the most studied BSM particles. Because they connect the lepton and hadron sectors, LQs are important ingredients in model-building exercises in various areas -- from dark matter to Higgs physics and many BSM scenarios, and leave interesting signatures at the current and future search facilities (see, e.g., Ref.~\cite{Cheung:2023gwm}). In these exercises, a collection like \textsc{TooLQit} can be handy as it can help determine the allowed parameter ranges and simulate various signatures in a uniform framework. The current version of \textsc{TooLQit} should be considered a proof-of-principle demonstration of the possibility of building a robust, modular, unified framework designed for LQ studies. We are working on the NLO-QCD \textsc{FeynRules} models, which we plan to include in the future. We plan to include support for all other single LQ models available in the literature in \textsc{CaLQ} as well as popular multi-LQ models such as $\tilde{R}_2+S_1/S_3$~\cite{Dorsner:2017wwn,Parashar:2022wrd,Dev:2024tto}, $S_1+S_3$~\cite{Bhaskar:2022vgk}, etc. We also plan to extend the coverage of \textsc{CaLQ} with LQs that interact with exotic fermions (like vectorlike quarks). Since the code is modular and the $\chi^2$ technique is generic, in the future, we like to enable custom model support where a user generates some specific processes to produce \texttt{.root} files, run simple codes (which we supply) on these files and import the output in \textsc{CaLQ} to calculate the limits.\footnote{Though not directly related to LQs, we note that \textsc{CaLQ} can be easily expanded to set limits on coupling-like parameters in BSM scenarios without LQs (e.g., Wilson coefficients in an effective-field-theory framework), which can modify the dilepton distributions.} We are also working on a likelihood-based alternative to the $\chi^2$ technique, which the user will be able to choose in future versions. We will also expand the data coverage of \textsc{CaLQ} by including the LHC data on monolepton plus missing energy (e.g., Refs.~\cite{CMS:2022ncp,ATLAS:2024tzc}) and mixed-flavours dilepton searches (e.g., Ref.~\cite{CMS:2022fsw}). In addition to this, we plan to include the limits from the latest direct searches (e.g., as listed in Ref.~\cite{Bhaskar:2023ftn}).

\acknowledgments 
\noindent 
T.M. acknowledges partial support from the SERB/ANRF, Government of India, through the Core Research Grant (CRG) No. CRG/2023/007031. R.S. acknowledges the PMRF from the Government of India. 

\bibliography{Leptoquark}

\end{document}